\documentclass[%
 reprint,
 amsmath,amssymb,
 aps
]{revtex4-1}
\pdfoutput=1
\usepackage{graphicx}% Include figure files
\usepackage{dcolumn}% Align table columns on decimal point
\usepackage{bm}% bold math
\usepackage{verbatim}
\usepackage{color}

\usepackage[utf8]{inputenc}

\begin{document}

\preprint{}[CCTP-2017-6, ITCP-IPP 2017/17]

\title{\huge Holographic Phonons}% Force line breaks with \\

\author{Lasma Alberte}%
 \email{lalberte@ictp.it}
\affiliation{Abdus Salam International Centre for Theoretical Physics (ICTP), Strada Costiera 11, 34151, Trieste, Italy.
}%

\author{Martin Ammon}%
 \email{martin.ammon@uni-jena.de}
\affiliation{Theoretisch-Physikalisches Institut, Friedrich-Schiller-Universit\"at Jena,
Max-Wien-Platz 1, D-07743 Jena, Germany.
}%

\author{Matteo Baggioli}%
 \email{mbaggioli@physics.uoc.gr}
\affiliation{Crete Center for Theoretical Physics, Institute for Theoretical and Computational Physics\\ Department of Physics, University of Crete, 71003
Heraklion, Greece.
}%

\author{Amadeo Jim\'enez-Alba}%
 \email{amadeo.jimenez.alba@uni-jena.de}
\affiliation{Theoretisch-Physikalisches Institut, Friedrich-Schiller-Universit\"at Jena,
Max-Wien-Platz 1, D-07743 Jena, Germany.
}%

\author{Oriol Pujol{\`a}s}%
 \email{pujolas@ifae.es}
\affiliation{Institut de F\'isica d'Altes Energies (IFAE), The Barcelona Institute of Science and Technology (BIST)\\
Campus UAB, 08193 Bellaterra, Barcelona.
}%

\begin{abstract}
We present a class of holographic massive gravity models that realize a spontaneous breaking of translational symmetry -- they  exhibit transverse phonon modes whose speed relates to the elastic shear modulus according to elasticity theory. Massive gravity theories thus emerge as versatile and convenient theories to model generic types of translational symmetry breaking: explicit, spontaneous and a mixture of both. The nature of the breaking is encoded in the radial dependence of the graviton mass. 
As an application of the model, we compute the temperature dependence of the shear modulus and find that it features a glass-like melting transition.

\end{abstract}

\pacs{Valid PACS appear here}
\maketitle

\section{Introduction}
In the last decade the gauge/gravity duality has proven to be an efficient tool to tackle condensed matter questions in the context of strongly coupled physics \cite{Hartnoll:2016apf,Ammon:2015wua,zaanen2015holographic}. Despite the various directions and applications pursued, a fundamental piece of the condensed matter phenomenology
is still missing in the holographic puzzle:
a concrete, simple and clear realization of phonons with standard properties as dictated by  elasticity theory, see \textit{e.g.} \cite{chaikin_lubensky_1995}. 
With the present letter we shall rectify this deficiency 
%intend to remedy this lack 
by presenting a class of simple holographic models featuring transverse phonons and elastic properties. \\

Recently there has been  significant progress towards including
translational symmetry breaking,  momentum dissipation and their consequences on transport in the context of holography \cite{Hartnoll:2012rj,Horowitz:2012ky} (see \cite{Hartnoll:2016apf} for a complete list of references). Within that framework, Massive Gravity (MG) stands out as a convenient and flexible  
%generic
%effective %theory in the 
gravity dual where the momentum relaxation time is set by the graviton mass $\tau^{-1}_{rel}\sim m_g^2$ \cite{Vegh:2013sk,Davison:2013jba,Blake:2013bqa}. The question regarding the nature of the translational symmetry breaking, \emph{i.e.} whether it occurs in a spontaneous or explicit manner, is however subtle. According to the holographic dictionary, the answer lies in the asymptotic UV behavior of the bulk fields breaking the translational invariance \cite{Klebanov:1999tb}. In the case of massive gravity this relates to the radial dependence of the graviton mass, which was shown  to admit a broad range of possible profiles $m_g(u)$ compatible with theoretical consistency  \cite{PhysRevLett.114.251602,Alberte:2015isw}.
%dependence $m_g(u)$ compatible with consistency.

A first evidence confirming this logic was presented in~\cite{Alberte:2017cch}, where gapped %and damped 
transverse phonons were identified, with the size of the gap being directly related to the asymptotic behaviour of the graviton mass. This suggests  a clear way to realize gapless phonons by ensuring a rapid enough decay of $m_g(u)$ towards the boundary. 

In this letter we consider a subclass of the holographic MG models introduced in \cite{PhysRevLett.114.251602,Alberte:2015isw} exhibiting such behaviour. We then demonstrate explicitly how it can attain a spontaneous symmetry breaking (SSB) of translations and provide a realization of massless phonons, \emph{i.e.} the corresponding Goldstone bosons. The resulting gapless modes show properties indentical to the transverse phonons in solids.
In particular, we find that their speed of propagation is in perfect agreement with the expectations from 
elasticity theory \cite{Landau7}.
To the best of our knowledge this is the first time that transverse phonons are realized within holography, with a sharp and clear relation to the elastic moduli as dictated by standard elasticity theory \footnote{The SSB of translations has been previously realized in holography in \cite{Nakamura:2009tf,Donos:2012wi,Donos:2013gda,Donos:2011bh,Andrade:2017cnc,Andrade:2017cnc,Jokela:2017ltu} as the dual of \textit{charge density waves} states \cite{RevModPhys.60.1129} where the Goldstone boson is identified with the so-called \textit{sliding mode}. Another realization of gapless transverse phonon modes was made in \cite{Esposito:2017qpj}, using a model with more dynamical ingredients -- even though a direct comparison to the elastic moduli is lacking. 
}.

Despite the fact that the physics of phonon excitations and elasticity in weakly coupled materials is well known, their holographic realization has remained absent for more than a decade. 
We believe that the present work will clarify how the phonons can be encoded in the holographic models. In addition, by the AdS/CFT dictionary, this will contribute to a better understanding of the role of phonons in strongly coupled materials.

\section{Holographic setup}
We consider generic \textit{solid} holographic massive gravity models \cite{PhysRevLett.114.251602,Alberte:2015isw}: 
\begin{equation}\label{S}
S\,=\, M_P^2\int d^4x \sqrt{-g}
\left[\frac{R}2+\frac{3}{\ell^2}- \, m^2 V(X)\,-\,\frac{1}{4}\,F^2\right]
\end{equation}
with $X \equiv \frac12 \, g^{\mu\nu} \,\partial_\mu \phi^I \partial_\nu \phi^I$ and $F^2=F_{\mu\nu}F^{\mu\nu}$. We study 4D AdS black brane geometries of the form:
\begin{equation}
\label{backg}
ds^2=\frac{\ell^2}{u^2} \left[\frac{du^2}{f(u)} -f(u)\,dt^2 + dx^2+dy^2\right] ~,
\end{equation}
where $u\in [0,u_h]$ is the radial holographic direction spanning from the boundary to the horizon, defined through $f(u_h)=0$, and $\ell$ is the AdS radius.\\

The $\phi^I$ scalars are the St\"uckelberg fields admitting a radially constant profile $\phi^I=x^I$ with $I=x,y$.
This is an exact solution of the system due to the shift symmetry. In the dual picture these fields represent scalar operators breaking the translational invariance because of the explicit dependence on the spatial coordinates. 
In this letter we shall consider benchmarks models of the type
\begin{equation}
\label{Xn}
V(X)=X^n~. 
\end{equation}
These are referred to as massive gravity theories because, among other reasons, the metric perturbations acquire a mass term given by $m^2_g(u) = 2m^2 X\, V'$ 
with the background value for $X=u^2/\ell^2$.
The absence of ghost and gradient instabilities enforces the conditions $V'>0$ and $c_{(bulk)}^2=1+\color{black}X\, V''/V'>0$ which constrain the power to satisfy $n>0$ \cite{PhysRevLett.114.251602}.\\

In the following we assume \emph{standard quantization}. This means that the near-boundary leading mode of the St\"uckelberg fields, $\phi^I_{(l)}$, sets the \emph{source} for the dual operator $\mathcal O_I$ breaking the translational invariance. The \emph{ expectation value} $\left\langle O_I\right\rangle$ is in turn set by the subleading mode $\phi^I_{(s)}$. 

For potentials of the type \eqref{Xn} the asymptotic expansion of the St\"uckelberg scalars close to the UV boundary at $u=0$ is given by:
\begin{equation}\label{bdy}
\phi^I\left(x^\mu\right)\,=\,\phi^I_{(0)}\left(t,x^i\right)\,+\,\phi^I_{(1)}\left(t,x^i\right)\,u^{5-2n}\,+\,\dots
\end{equation}
where $\phi^I_{(0)}\left(t,x^i\right)=x^I\neq 0$ on the bulk solution.
Depending on the value of $n$ one can then distinguish two cases. 
If $n<5/2$ then $\phi^I_{(0)}$ is the leading term in the near-boundary expansion and corresponds to the source, \emph{i.e.} $\phi^I_{(l)}=x^I$. As a consequence, the dual QFT contains an explicit breaking term which gives rise to a finite relaxation time $\tau_{rel}^{-1}\sim m_g^2$ for the momentum operator \cite{Davison:2013jba}. This is the case for all the potentials $V(X)$ that have so far been considered in the literature \cite{PhysRevLett.114.251602,Alberte:2015isw,Alberte:2017cch}.

The main observation is that if instead one considers a potential of the form $V(X)=X^n$ with a sufficiently large $n>5/2$, the mode $\phi^I_{(0)}$  becomes subleading in the boundary expansion \eqref{bdy}. Hence, for $n>5/2$ the solution $\phi^I=x^I$ for the scalar bulk fields gives rise to an expectation value $\langle \mathcal{O}_I\rangle \neq 0$ for its dual operator while its source vanishes, leading to the SSB pattern \footnote{We thank Blaise Gouteraux for suggesting this interpretation. See also  \cite{Amoretti:2016bxs} for a detailed analysis on the nature of the translational symmetry breaking patterns in QFT and holography.
}. Intuitively such a condition corresponds to demanding that the radially dependent graviton mass $m_g(u)$ is large at the horizon and quickly vanishes at the boundary, as already suggested in \cite{PhysRevLett.114.251602,Alberte:2017cch}.\\

Let us now focus on an important feature of the benchmark models \eqref{Xn}. For $n\neq1$ -- our main focus in this work -- the kinetic term for the St\"uckelberg fields $\phi^I$ is non-canonical: they do not have a quadratic action. This implies that their quantization is at best non-standard, and thus the theory is strongly coupled on and near the trivial solution $\phi^I=$ ~const. For this reason, the interpretation of this classical solution as a valid quantum vacuum is quite dubious. 
Instead, for the non-trivial solution, $\phi^I=x^I$, the conclusion changes in a radical way.

The main point is simply that around a non-trivial solution the St\"uckelberg fields do acquire a standard quadratic kinetic term, at least in part of the geometry. To illustrate this, it suffices to consider the St\"uckelberg fields in the transverse sector. Separating them into background and perturbations as $\phi^I= x^I + \pi^I$, with $\partial_i\pi^i=0 $, one can easily expand the Lagrangian in powers of $\pi$'s to find
$$
- M_P^2 m^2 \int d^4x\,\sqrt{-g} \left\{ V'(\bar X) X_\pi+\frac12
 V''(\bar X) X_\pi^2 + \dots \right\}
$$
where $\bar X$ denotes the background value, $u^2/\ell^2$, and $X_\pi \equiv \frac12 \, g^{\mu\nu} \,\partial_\mu \pi^I \partial_\nu \pi^I$. One can estimate the strong coupling scale in this sector by going to canonical normalization $\pi_{c} = M_P m \sqrt{V'} \; \pi$ and finding the scale suppressing the dimension-8 operator $(\partial \pi_c)^4 /\Lambda(u)^4$. The strong coupling scale $\Lambda(u)$ is actually radial dependent and is given by
$\Lambda(u)^{4} \sim \frac{V'^2}{V''} M_P^2 m^2$. For our benchmark models \eqref{Xn} with generic $n$ this translates into
\begin{equation}\label{Lambda}
\Lambda(u) \sim (m\,M_P)^{1/2} 
\left(\frac{u}{\ell}\right)^{n/2}\,.
\end{equation}
Thus, for a fixed mass parameter $m^2$ the transverse field perturbations $\pi^I$ become strongly coupled above a different energy scale $\Lambda(u)$ for each value of the radial distance~$u$. 

Since $\Lambda(u)$ asymptotically vanishes towards the AdS boundary located at $u=0$, then in practice our EFT in the bulk is tractable only down to a certain radius, for $u > u_*$. In the dual CFT the scale $u_*$ clearly corresponds to some UV cutoff. This is a very welcome feature since we do expect the strength of phonon self-interactions to increase towards high energies, as they  do in  weakly coupled materials, see \emph{e.g.} \cite{Leutwyler:1996er}. The scale $u_*$ is thus naturally identified with the lattice spacing scale $a$, setting an upper cutoff to the phonons frequency. In physical terms this cutoff relates to the fact that the phononic vibrational modes cannot be excited above the so-called Debye temperature \cite{2007physics...3001G}. Therefore it is neither surprising nor problematic to have a UV cutoff in our gravitational theory; on the contrary, it is an important physical property which makes these models more realistic. How small $u_*$ is depends on how small we can tolerate the St\"uckelberg strong coupling scale, $\Lambda_*$, anywhere in the bulk. Notice that this is a new parameter in the model, independent from the parameters that appear in \eqref{S}\footnote{This new parameter does not appear in models with a standard kinetic term near $X=0$, \emph{i.e.}, $V(X)=X+X^2+\dots$ because there $\Lambda(u)$ does not asymptote to zero}.

There are two basic and obvious constraints in choosing $\Lambda_*$. First, $\Lambda_*$ must be bigger than the typical gradients, that is, $\Lambda_* \gg 1/\ell$. Second, in order to still be able to read off the holographic correlators from the decay modes of the bulk fields, we also need  the UV cutoff $u_*$ to be close to the AdS boundary, so that $u_* \ll \ell$. In other words, we must ensure that the ratio $u_*/\ell$ is sufficiently small. 
 
Thus we can write
$\Lambda_* = (m\,M_P)^{1/2} 
(u_*/\ell)^{n/2}$ and think of $u_*/\ell$ as a fixed small number. Requiring that $\Lambda_* \ell \gtrsim1$ then gives 
$$
M_P\ell \gtrsim \frac{1}{m \ell} \left(\frac{u_*}{\ell}\right)^{-n}  \, .
$$
Note that $M_P \ell$ has to be large for the employed semi-classical treatment of the gravitational side to be valid. Hence, for any given small $u_*/\ell$ and any $n$, we can satisfy the two constraints with $m \ell$ of order one. Once  these conditions are met, then these constructions allow to model in a controlled way the physics of phonons in critical/conformal solids.

\begin{figure}[htp]
\begin{center}
\includegraphics[width=0.48\textwidth]{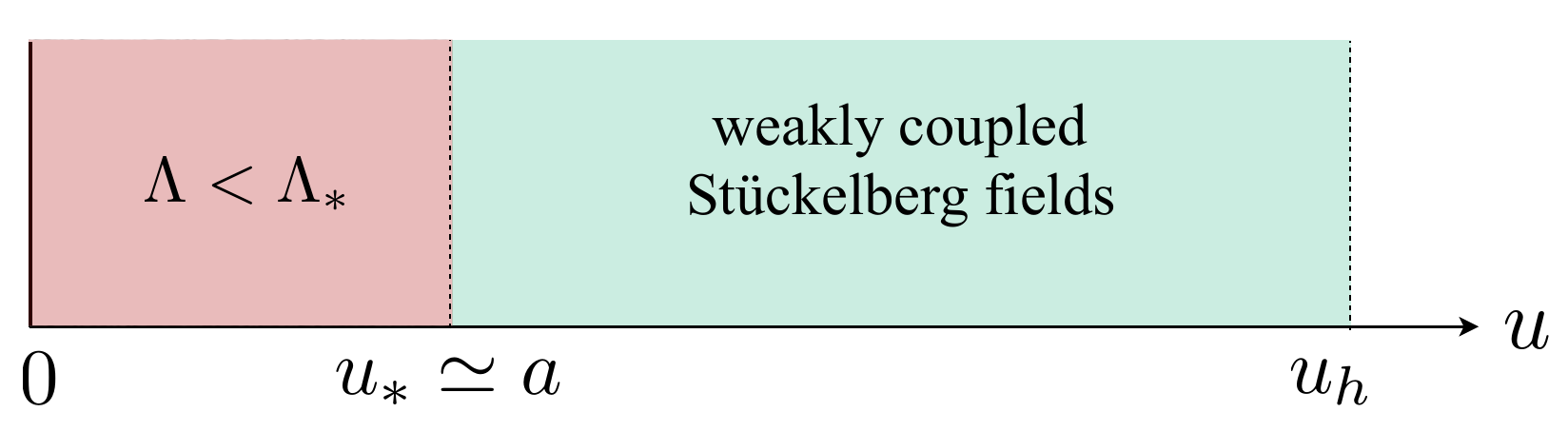}%&
 \caption{A sketch of the validity of the EFT in the bulk and its dual interpretation.}
 \label{fig:sketch}
\end{center}
\end{figure}

\section{Results}
\subsection{Phonons and elasticity}
In solids translational invariance is spontaneously broken. The corresponding Goldstone bosons -- the phonons -- play a crucial role in the description of the low energy physics and the elastic properties of the materials. Their dynamics can be entirely captured via effective field theory methods \cite{Leutwyler:1996er,Nicolis:2015sra}. Depending on the direction of propagation with respect to the deformation of the medium they can be classified into longitudinal and transverse phonons; in this letter we shall focus on the latter. The presence of propagating transverse phonons, also called the \textit{shear sound}, is a characteristic property of solids and provides a clear physical distinction from fluids.\\

The dispersion relation for the transverse phonons takes the simple form \cite{chaikin_lubensky_1995}:
\begin{equation}\label{disprel}
\omega\,=\,c_T\,k\,-\,i\,D\,k^2\,,
\end{equation}
where $c_T$ is the speed of propagation and $D$ is the momentum diffusion constant, proportional to the finite viscosity $\eta$ of the medium. In the absence of explicit breaking, neither mass gap nor damping is present \footnote{This is opposed to the case of mass potentials $V(X)=X^{n}$ with $ n<5/2$, studied in \cite{Davison:2013jba,Davison:2014lua,Alberte:2017cch}.}. In relativistic systems, the velocity is set to be \cite{PhysRevA.6.2401,1980PhRvB..22.2514Z,1963AnPhy..24..419K,chaikin_lubensky_1995}:
\begin{equation}\label{speedformula}
c_T^2\,=\,\frac{G}{\chi_{PP}}\,,
\end{equation}
where $G$ is the \textit{shear elastic modulus} and $\chi_{PP}$ is the momentum susceptibility. Both of these quantities can be extracted via the Kubo formulas from the $T_{xy}$ (shear stress) and $T_{tx}$ (momentum) retarded correlators as follows:
\begin{align}
G\,\equiv\, \mathcal{G}^{\textrm{(R)}}_{T_{xy}T_{xy}}\,\bigg|_{\omega,k=0}\,,\hspace{1cm}\chi_{PP}\,\equiv \, \mathcal{G}^{\textrm{(R)}}_{T_{tx}T_{tx}}\,\bigg|_{\omega,k=0}\,.
\end{align}
%We can show (see Appendix~\ref{SUPPL}) that
The momentum susceptibility coincides with \cite{Hartnoll:2016apf}:
\begin{equation}
\chi_{PP}\,=\,T_{00}\,+\,T_{xx}\,=\,\epsilon\,+\,\mathcal{P}\,=\,\frac{3}{2}\,\epsilon\,,
\end{equation}
where $\epsilon$ is the energy density and $\mathcal{P}$ is the mechanical pressure \footnote{Due to the presence of a finite shear modulus $G$ the mechanical pressure, $\mathcal{P}\equiv T_{xx}$, and the thermodynamic pressure, $p\equiv -\Omega/\mathcal V$, are not equivalent.}.\\
\begin{figure}[htp]
\begin{center}
\includegraphics[width=0.48\textwidth]{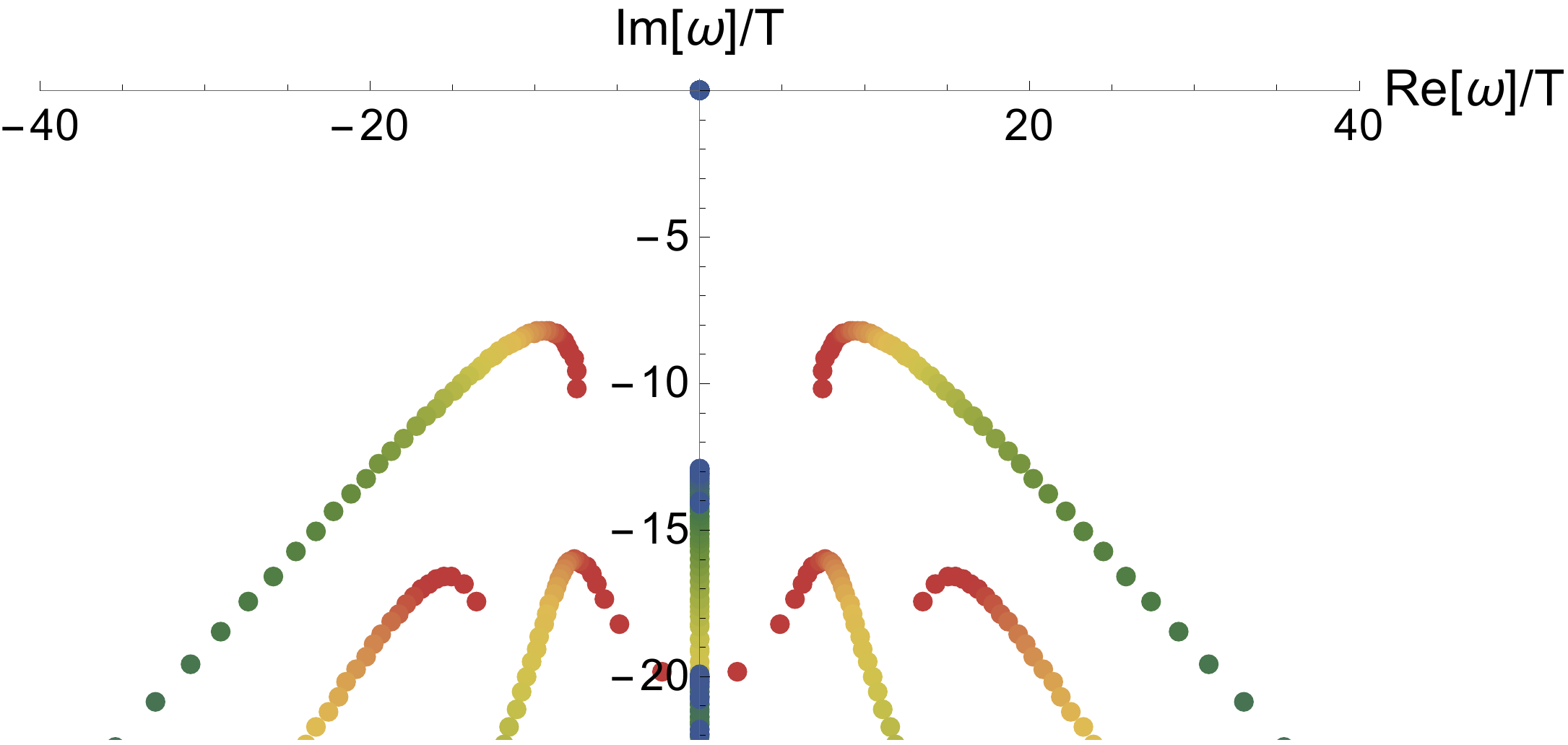}
 \caption{The quasi-normal spectrum for $V(X)=X^5$ at zero momentum for $T/m\in[0.01,1.05]$ (blue--red). 
 }
 \label{fig:QNM}
\end{center}
\end{figure}\\
In order to confirm the presence of transverse phonons and verify Eq.~\eqref{speedformula}, we find the spectrum of the quasi-normal modes (QNMs) of the system in the transverse sector. Here we have done it explicitly for potentials of the type $V(X)=X^n$ for several values of $n\in[3-8]$ including also non-integer values \footnote{We have also checked the linear superpositions of the type $V(X)=\sum c_\nu X^\nu$ with $\nu>5/2$; the qualitatitve results remain unchanged.}. For concreteness, some of the following plots only show specific realizations; nevertheless the qualitative conclusions drawn from the data are the same in all cases. 

In Fig.~\ref{fig:QNM} we show the spectrum of QNMs at zero momentum and different temperatures for $n=5$. We find a QNM located at zero frequency, corresponding to a gapless quasi-particle, the putative phonon in our holographic model. Moreover, we note that for any temperature the next QNM is already highly damped.
\begin{figure}[htp]
\begin{center}
\includegraphics[width=0.23\textwidth]{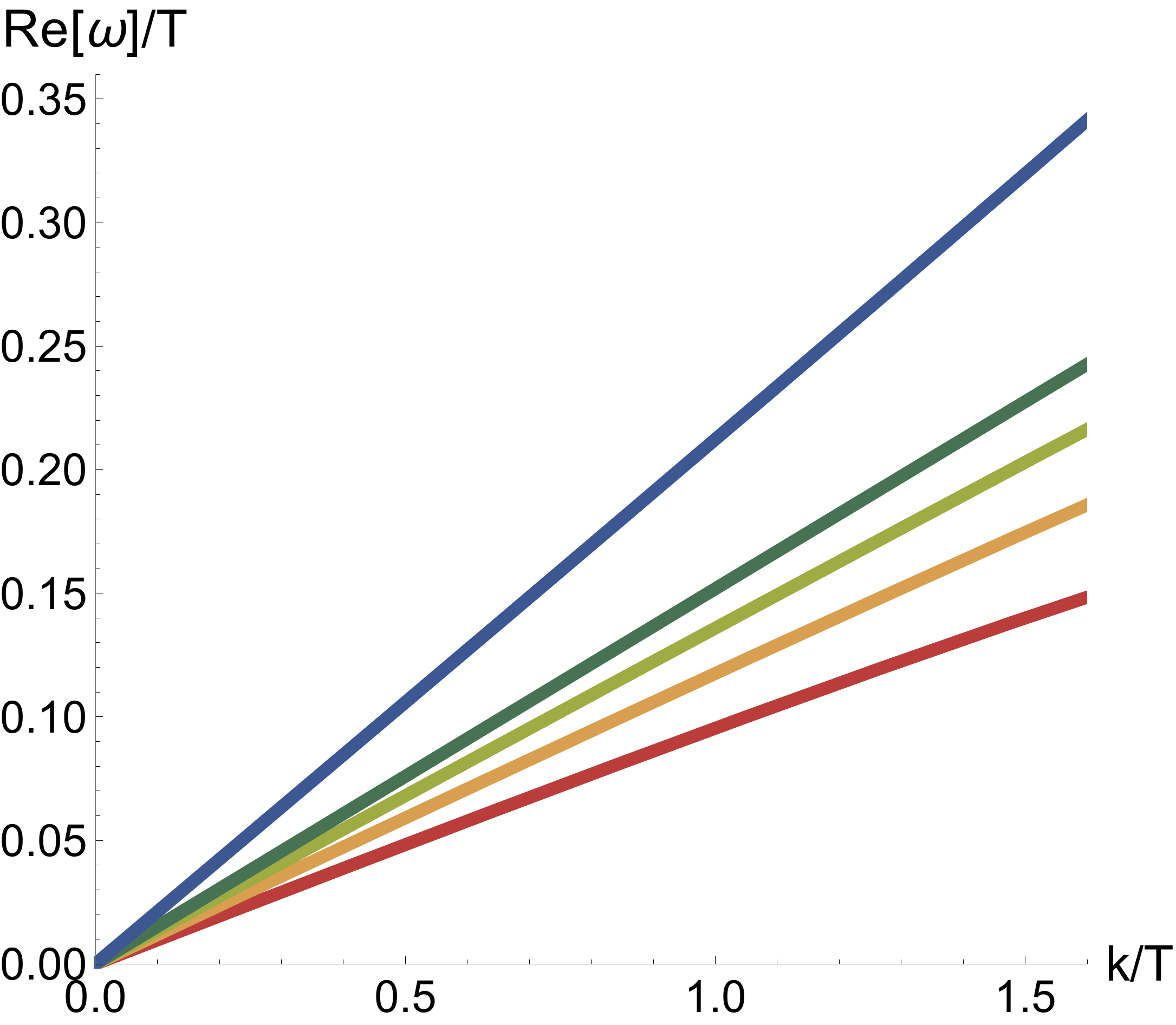}
\includegraphics[width=0.23\textwidth]{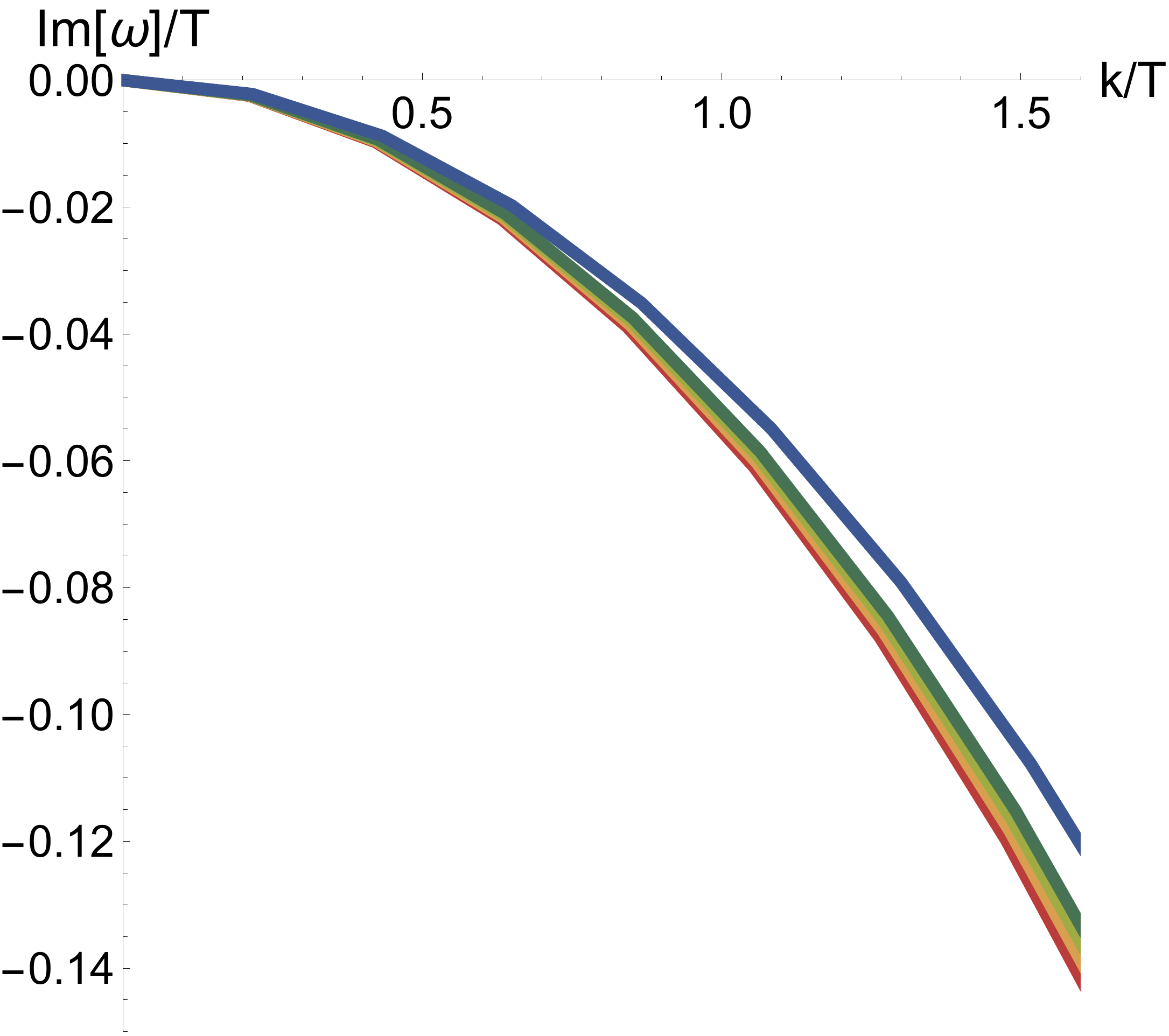}
 \caption{Real and imaginary parts of the frequency of the lowest QNM for $V(X)=X^5$ for $T/m\in[0.7-1.7]$ %$m/T=(0.6-1.4)$ (red--blue).
 (blue--red).}\label{fig:im}
\end{center}
\end{figure}

\begin{figure}[htp]
\begin{center}
\includegraphics[width=0.48\textwidth]{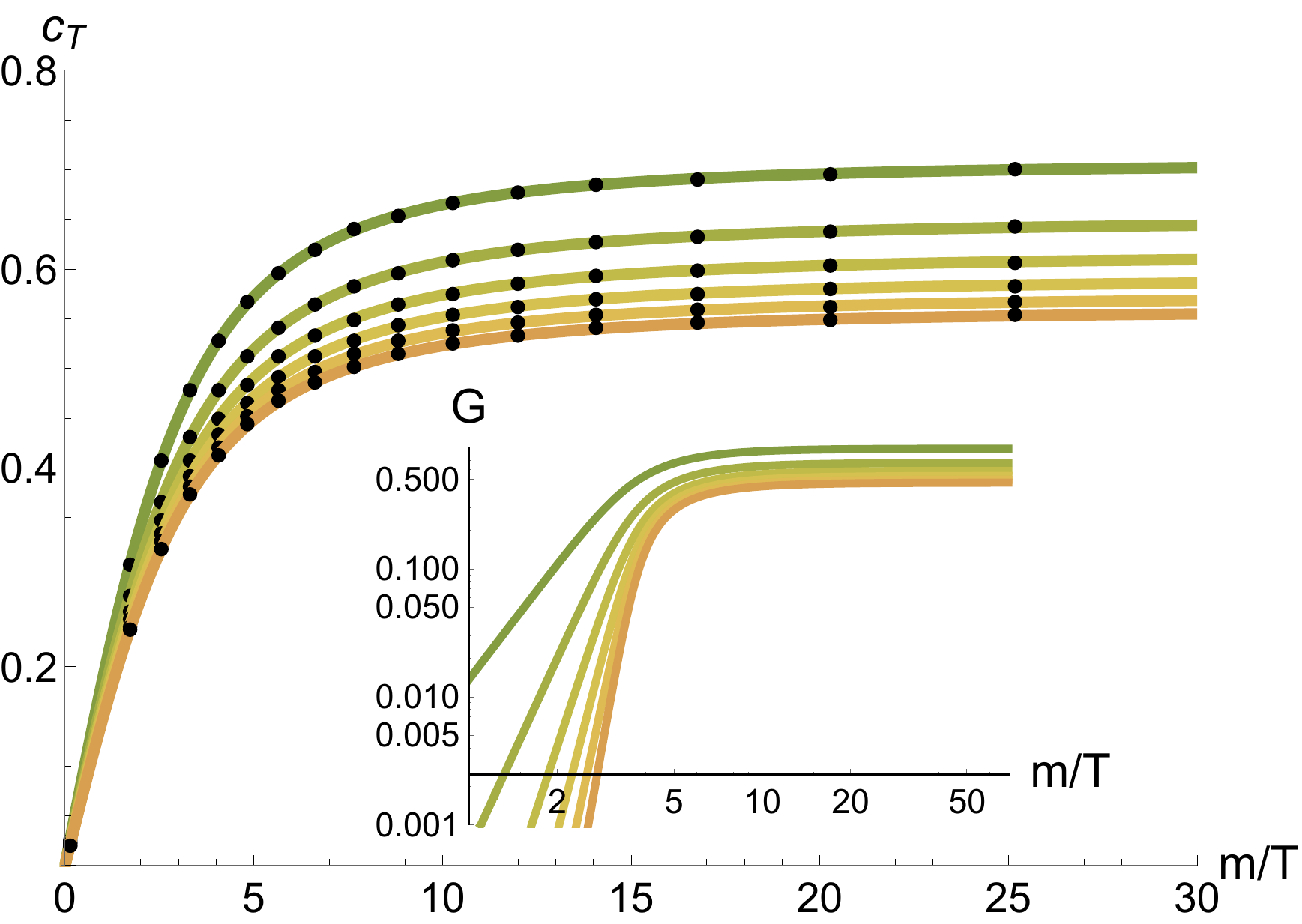}
 \caption{Comparison of the velocity extracted from the QNMs (black dots) and the velocity computed from the elasticity (solid lines) for $n\in[3-8]$ (green--orange). Inset: The dependence of the shear elastic modulus on $m/T$ in log-log scale. The fall-off at large temperatures $G\sim m^2T^{3-2n}$ is evident. See Appendix~\ref{SUPPL} for more details. 
 }
 \label{fig:comparison}
\end{center}
\end{figure}

Next, we analyze the QNM spectrum at finite momentum $k$. In Fig.~\ref{fig:im} we show the behavior of both its real and imaginary parts. As evident from Fig.~\ref{fig:im}, this quasi-normal mode satisfies the expected dispersion relation of Eq.~\eqref{disprel} \footnote{We fitted the data to $\omega= C_1k^{\alpha_1}+i\,C_2k^{\alpha_2}$. In all cases we found similar values:  $\alpha_1=1.00001\pm4\cdot10^{-5}$ and $\alpha_2=2.00002\pm1\cdot10^{-5}$.}. Hence it is neither attenuated nor gapped. We note that the speed $c_T$ decreases with increasing temperature (see Fig.~\ref{fig:comparison}) while the diffusion constant $D$ increases with $T/m$.  In particular, at $T/m=0$ the diffusion constant and the viscosity vanish $D=\eta=0$ (see also \cite{Hartnoll:2016tri,Alberte:2016xja}) and the elastic modulus $G$ is maximal. In turn, at $T/m\gg1$ the viscosity is maximal and the elastic modulus is zero. In other words, the physics interpolates from a solid behaviour at zero temperature to a fluid behaviour at high temperatures in a continuous way, exhibiting viscoelastic features in the intermediate temperature range. This feature is qualitatively similar to a glassy transition typical to viscoelastic materials \cite{RevModPhys.78.953,RevModPhys.83.587,2009PhR...476...51C}, as shown in the inset of Fig.~\ref{fig:comparison} and explained in more detail in Appendix~\ref{SUPPL}.\\

Crucially, as shown in Fig.~\ref{fig:comparison}, the sound speed of the transverse phonons extracted from the quasi-normal mode analysis is in perfect agreement with the expectations from the elastic theory given in Eq.~\eqref{speedformula}. %This is our main result.
We note that the value of the velocity in the zero-temperature limit is always subluminal, but is \emph{not} universal, contrary to an earlier probe-limit claim of~\cite{Esposito:2017qpj} Moreover, we find that for $n\geq 3$ the sound speed satisfies the bound $c_T^2\leq 1/2$ arising for conformal solids~\cite{Esposito:2017qpj}.

\subsection{Conductivity and viscosity}
By introducing a finite charge density $\rho$ we are able to analyze also the electric optical conductivity $\sigma(\omega)$ of our system.
In the presence of only SSB its low frequency expansion is expected to be given by \cite{Hartnoll:2016apf}:
\begin{equation}\label{sigma}
\sigma(\omega)\,=\,\sigma_Q\,+\,\frac{\rho^2}{\chi_{PP}}\,\left(\delta(\omega)\,+\,\frac{i}{\omega}\right)\,,
\end{equation}
where $\sigma_Q$ is the so-called \textit{incoherent conductivity} (see \cite{Davison:2014lua,Davison:2015taa}). Notice that because of the absence of a finite momentum relaxation time, $\tau_{rel}$, or equivalently of an explicit breaking mechanism, the DC conductivity is infinite.\\

\begin{figure}[htp]
\begin{center}
\vspace{0.15cm}
\includegraphics[width=0.48\textwidth]{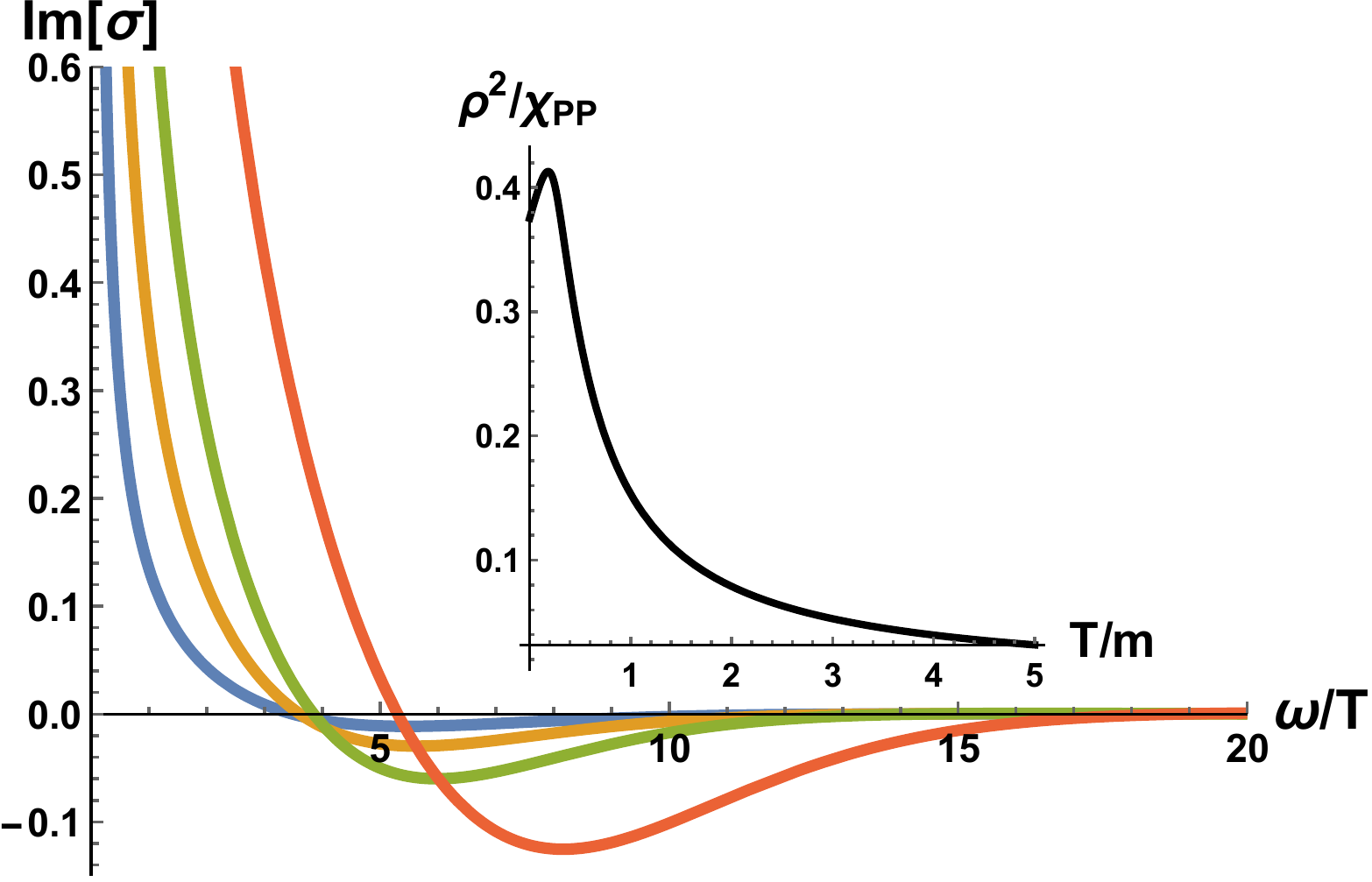}
 \caption{\label{fig:conductivity}Imaginary part of $\sigma(\omega)$ for $V(X)=X^5$ at $\mu=1$ and different $T/m=1,0.6,0.4,0.2$. In the inset the Drude weight, $\rho^2/\chi_{PP}$, is shown. %The agreement between the numerics and the formula is perfect.
 }
\end{center}
\end{figure}

We have computed the optical conductivity of our class of models via the Kubo formula: 
\begin{equation}
\sigma(\omega)\,=\,\frac{1}{i\,\omega}\,\mathcal{G}^{\rm (R)}_{JJ}(\omega)\,\bigg|_{k=0}\,.
\end{equation}
As shown in Fig.~\ref{fig:conductivity}, the low frequency behaviour of the conductivity agrees with the form presented in \eqref{sigma}; also the Drude weight is in perfect agreement with the hydrodynamics expectations.
This allows us to confidently claim that our setup \eqref{S} does indeed represent a dual of the SSB of translational invariance, together with all its physical manifestations.

Finally, let us comment on the momentum diffusion constant $D$ appearing in the dispersion relation \eqref{disprel} and related to the \textit{hydrodynamical} viscosity $\eta$ by $D=\eta/\chi_{PP}\,$. We find that $\eta$ does not agree with the value extracted via the Kubo formula: 
\begin{equation}
\eta^*\,=\,-\,\lim_{\omega\to 0}\,\frac{1}{\omega}\,\textrm{Im}\,\mathcal{G}^{\rm (R)}_{T_{xy}T_{xy}}(k=0)\,.%\bigg|_{\omega,k=0},
\end{equation}
This was already noticed for the explicit breaking case with $n=1$ in \cite{Burikham:2016roo,Ciobanu:2017fef}. Here we confirm this disagreement between the two viscosities for generic values of $n$ by a direct numerical computation of the shear mode dispersion relation.\\

\section{Conclusions}
We present a simple holographic gravity dual for phonons in strongly coupled materials, whose properties are in perfect agreement with elasticity theory.
Our results open a new window for the study of strongly coupled solids  via the AdS/CFT methods. In the process, we also sharpen and quantify the connection between elastic theory and massive gravity \cite{Vegh:2013sk,PhysRevLett.114.251602,Alberte:2015isw,Beekman:2016szb}.

The future possibilities are diverse. 
One direction of clear interest is to compute and characterize the %elastic and 
viscoelastic response of these models in more detail. This will clarify the connection with the known glassy melting transitions \cite{RevModPhys.78.953,RevModPhys.83.587,2009PhR...476...51C}. In this regard it seems relevant to study the response under time-dependent stresses as it could shed light on further signatures typical for %glassy/
amorphous and viscoelastic materials like slow relaxation and aging \cite{0034-4885-79-6-066504} (see \emph{e.g.} \cite{Anninos:2013mfa} for previous works).

In addition, it would be compelling to elucidate any possible relation to the so-called \textit{quantum critical elasticity}~\cite{2015PhRvL.115b5703Z,2015EPJST.224.1021Z}.

Finally these methods can be used to model the optical transport properties of the strange/bad metals in terms of  the interplay between the explicit and spontaneous breaking of translations \cite{Delacretaz:2016ivq,Delacretaz:2017zxd}.\\[0.1cm]

%We plan to continue in these directions in the near future.\\[0.1cm]

\section*{Acknowledgements}
We thank A. Amoretti, D. Arean, A. Beekman, S. Grozdanov, S. Hartnoll, K.Y. Kim, A. Krikun, J. Leiber, D. Musso, N. Obers, C. Pantelidou, N. Poovuttikul, S.J. Sin, J. Zaanen and in particular B. Gouteraux for useful discussions and comments about this work and the topics considered. We are grateful to J. Zaanen for reading a preliminary version of this letter and providing insightful comments.\\
MB is supported in part by the Advanced ERC grant SM-grav, No 669288. AJ acknowledges financial support by Deutsche Forschungsgemeinschaft (DFG) GRK 1523/2. OP acknowledges support by the Spanish Ministry MEC under grant FPA2014-55613-P and the Severo Ochoa excellence program of MINECO (grant SO-2012-0234, SEV-2016-
0588), as well as by the Generalitat de Catalunya under grant 2014-SGR-1450. MB would like to
thank the Nordic Institute for Theoretical Physics (NORDITA) and the organizers of the ''Many-Body Quantum Chaos, Bad Metals and Holography'' workshop for the hospitality during the completion of this work.
MB would also like to ``thank'' the University General Hospital of Heraklion Pagni for the long hospitality during the completion of this letter.\\[0.05cm]
\textbf{Note added.} 
While this letter was being completed we became aware of upcoming works discussing similar issues \cite{NickAppear,BlaiseAppear}. %We are grateful to the authors for sharing their preliminary results.
\bibliographystyle{apsrev4-1}
\bibliography{phonons}

\begin{thebibliography}{56}%
\makeatletter
\providecommand \@ifxundefined [1]{%
 \@ifx{#1\undefined}
}%
\providecommand \@ifnum [1]{%
 \ifnum #1\expandafter \@firstoftwo
 \else \expandafter \@secondoftwo
 \fi
}%
\providecommand \@ifx [1]{%
 \ifx #1\expandafter \@firstoftwo
 \else \expandafter \@secondoftwo
 \fi
}%
\providecommand \natexlab [1]{#1}%
\providecommand \enquote  [1]{``#1''}%
\providecommand \bibnamefont  [1]{#1}%
\providecommand \bibfnamefont [1]{#1}%
\providecommand \citenamefont [1]{#1}%
\providecommand \href@noop [0]{\@secondoftwo}%
\providecommand \href [0]{\begingroup \@sanitize@url \@href}%
\providecommand \@href[1]{\@@startlink{#1}\@@href}%
\providecommand \@@href[1]{\endgroup#1\@@endlink}%
\providecommand \@sanitize@url [0]{\catcode `\\12\catcode `\$12\catcode
  `\&12\catcode `\#12\catcode `\^12\catcode `\_12\catcode `\%12\relax}%
\providecommand \@@startlink[1]{}%
\providecommand \@@endlink[0]{}%
\providecommand \url  [0]{\begingroup\@sanitize@url \@url }%
\providecommand \@url [1]{\endgroup\@href {#1}{\urlprefix }}%
\providecommand \urlprefix  [0]{URL }%
\providecommand \Eprint [0]{\href }%
\providecommand \doibase [0]{http://dx.doi.org/}%
\providecommand \selectlanguage [0]{\@gobble}%
\providecommand \bibinfo  [0]{\@secondoftwo}%
\providecommand \bibfield  [0]{\@secondoftwo}%
\providecommand \translation [1]{[#1]}%
\providecommand \BibitemOpen [0]{}%
\providecommand \bibitemStop [0]{}%
\providecommand \bibitemNoStop [0]{.\EOS\space}%
\providecommand \EOS [0]{\spacefactor3000\relax}%
\providecommand \BibitemShut  [1]{\csname bibitem#1\endcsname}%
\let\auto@bib@innerbib\@empty
%</preamble>
\bibitem [{\citenamefont {Hartnoll}\ \emph
  {et~al.}(2016{\natexlab{a}})\citenamefont {Hartnoll}, \citenamefont {Lucas},\
  and\ \citenamefont {Sachdev}}]{Hartnoll:2016apf}%
  \BibitemOpen
  \bibfield  {author} {\bibinfo {author} {\bibfnamefont {S.~A.}\ \bibnamefont
  {Hartnoll}}, \bibinfo {author} {\bibfnamefont {A.}~\bibnamefont {Lucas}}, \
  and\ \bibinfo {author} {\bibfnamefont {S.}~\bibnamefont {Sachdev}},\
  }\href@noop {} {\  (\bibinfo {year} {2016}{\natexlab{a}})},\ \Eprint
  {http://arxiv.org/abs/1612.07324} {arXiv:1612.07324 [hep-th]} \BibitemShut
  {NoStop}%
%%CITATION = ARXIV:1612.07324;%%
\bibitem [{\citenamefont {Ammon}\ and\ \citenamefont
  {Erdmenger}(2015)}]{Ammon:2015wua}%
  \BibitemOpen
  \bibfield  {author} {\bibinfo {author} {\bibfnamefont {M.}~\bibnamefont
  {Ammon}}\ and\ \bibinfo {author} {\bibfnamefont {J.}~\bibnamefont
  {Erdmenger}},\ }\href
  {http://www.cambridge.org/de/academic/subjects/physics/theoretical-physics-and-mathematical-physics/gaugegravity-duality-foundations-and-applications}
  {\emph {\bibinfo {title} {{Gauge/gravity duality}}}}\ (\bibinfo  {publisher}
  {Cambridge University Press},\ \bibinfo {year} {2015})\BibitemShut {NoStop}%
%%CITATION = INSPIRE-1376202;%%
\bibitem [{\citenamefont {Zaanen}\ \emph {et~al.}(2015)\citenamefont {Zaanen},
  \citenamefont {Liu}, \citenamefont {Sun},\ and\ \citenamefont
  {Schalm}}]{zaanen2015holographic}%
  \BibitemOpen
  \bibfield  {author} {\bibinfo {author} {\bibfnamefont {J.}~\bibnamefont
  {Zaanen}}, \bibinfo {author} {\bibfnamefont {Y.}~\bibnamefont {Liu}},
  \bibinfo {author} {\bibfnamefont {Y.}~\bibnamefont {Sun}}, \ and\ \bibinfo
  {author} {\bibfnamefont {K.}~\bibnamefont {Schalm}},\ }\href
  {https://books.google.gr/books?id=MrRynQAACAAJ} {\emph {\bibinfo {title}
  {Holographic Duality in Condensed Matter Physics}}}\ (\bibinfo  {publisher}
  {Cambridge University Press},\ \bibinfo {year} {2015})\BibitemShut {NoStop}%
\bibitem [{\citenamefont {Chaikin}\ and\ \citenamefont
  {Lubensky}(1995)}]{chaikin_lubensky_1995}%
  \BibitemOpen
  \bibfield  {author} {\bibinfo {author} {\bibfnamefont {P.~M.}\ \bibnamefont
  {Chaikin}}\ and\ \bibinfo {author} {\bibfnamefont {T.~C.}\ \bibnamefont
  {Lubensky}},\ }\href {\doibase 10.1017/CBO9780511813467} {\emph {\bibinfo
  {title} {Principles of Condensed Matter Physics}}}\ (\bibinfo  {publisher}
  {Cambridge University Press},\ \bibinfo {year} {1995})\BibitemShut {NoStop}%
\bibitem [{\citenamefont {Hartnoll}\ and\ \citenamefont
  {Hofman}(2012)}]{Hartnoll:2012rj}%
  \BibitemOpen
  \bibfield  {author} {\bibinfo {author} {\bibfnamefont {S.~A.}\ \bibnamefont
  {Hartnoll}}\ and\ \bibinfo {author} {\bibfnamefont {D.~M.}\ \bibnamefont
  {Hofman}},\ }\href {\doibase 10.1103/PhysRevLett.108.241601} {\bibfield
  {journal} {\bibinfo  {journal} {Phys. Rev. Lett.}\ }\textbf {\bibinfo
  {volume} {108}},\ \bibinfo {pages} {241601} (\bibinfo {year} {2012})},\
  \Eprint {http://arxiv.org/abs/1201.3917} {arXiv:1201.3917 [hep-th]}
  \BibitemShut {NoStop}%
%%CITATION = ARXIV:1201.3917;%%
\bibitem [{\citenamefont {Horowitz}\ \emph {et~al.}(2012)\citenamefont
  {Horowitz}, \citenamefont {Santos},\ and\ \citenamefont
  {Tong}}]{Horowitz:2012ky}%
  \BibitemOpen
  \bibfield  {author} {\bibinfo {author} {\bibfnamefont {G.~T.}\ \bibnamefont
  {Horowitz}}, \bibinfo {author} {\bibfnamefont {J.~E.}\ \bibnamefont
  {Santos}}, \ and\ \bibinfo {author} {\bibfnamefont {D.}~\bibnamefont
  {Tong}},\ }\href {\doibase 10.1007/JHEP07(2012)168} {\bibfield  {journal}
  {\bibinfo  {journal} {JHEP}\ }\textbf {\bibinfo {volume} {07}},\ \bibinfo
  {pages} {168} (\bibinfo {year} {2012})},\ \Eprint
  {http://arxiv.org/abs/1204.0519} {arXiv:1204.0519 [hep-th]} \BibitemShut
  {NoStop}%
%%CITATION = ARXIV:1204.0519;%%
\bibitem [{\citenamefont {Vegh}(2013)}]{Vegh:2013sk}%
  \BibitemOpen
  \bibfield  {author} {\bibinfo {author} {\bibfnamefont {D.}~\bibnamefont
  {Vegh}},\ }\href@noop {} {\  (\bibinfo {year} {2013})},\ \Eprint
  {http://arxiv.org/abs/1301.0537} {arXiv:1301.0537 [hep-th]} \BibitemShut
  {NoStop}%
%%CITATION = ARXIV:1301.0537;%%
\bibitem [{\citenamefont {Davison}(2013)}]{Davison:2013jba}%
  \BibitemOpen
  \bibfield  {author} {\bibinfo {author} {\bibfnamefont {R.~A.}\ \bibnamefont
  {Davison}},\ }\href {\doibase 10.1103/PhysRevD.88.086003} {\bibfield
  {journal} {\bibinfo  {journal} {Phys. Rev.}\ }\textbf {\bibinfo {volume}
  {D88}},\ \bibinfo {pages} {086003} (\bibinfo {year} {2013})},\ \Eprint
  {http://arxiv.org/abs/1306.5792} {arXiv:1306.5792 [hep-th]} \BibitemShut
  {NoStop}%
%%CITATION = ARXIV:1306.5792;%%
\bibitem [{\citenamefont {Blake}\ and\ \citenamefont
  {Tong}(2013)}]{Blake:2013bqa}%
  \BibitemOpen
  \bibfield  {author} {\bibinfo {author} {\bibfnamefont {M.}~\bibnamefont
  {Blake}}\ and\ \bibinfo {author} {\bibfnamefont {D.}~\bibnamefont {Tong}},\
  }\href {\doibase 10.1103/PhysRevD.88.106004} {\bibfield  {journal} {\bibinfo
  {journal} {Phys. Rev.}\ }\textbf {\bibinfo {volume} {D88}},\ \bibinfo {pages}
  {106004} (\bibinfo {year} {2013})},\ \Eprint {http://arxiv.org/abs/1308.4970}
  {arXiv:1308.4970 [hep-th]} \BibitemShut {NoStop}%
%%CITATION = ARXIV:1308.4970;%%
\bibitem [{\citenamefont {Klebanov}\ and\ \citenamefont
  {Witten}(1999)}]{Klebanov:1999tb}%
  \BibitemOpen
  \bibfield  {author} {\bibinfo {author} {\bibfnamefont {I.~R.}\ \bibnamefont
  {Klebanov}}\ and\ \bibinfo {author} {\bibfnamefont {E.}~\bibnamefont
  {Witten}},\ }\href {\doibase 10.1016/S0550-3213(99)00387-9} {\bibfield
  {journal} {\bibinfo  {journal} {Nucl. Phys.}\ }\textbf {\bibinfo {volume}
  {B556}},\ \bibinfo {pages} {89} (\bibinfo {year} {1999})},\ \Eprint
  {http://arxiv.org/abs/hep-th/9905104} {arXiv:hep-th/9905104 [hep-th]}
  \BibitemShut {NoStop}%
%%CITATION = HEP-TH/9905104;%%
\bibitem [{\citenamefont {Baggioli}\ and\ \citenamefont
  {Pujol\`as}(2015)}]{PhysRevLett.114.251602}%
  \BibitemOpen
  \bibfield  {author} {\bibinfo {author} {\bibfnamefont {M.}~\bibnamefont
  {Baggioli}}\ and\ \bibinfo {author} {\bibfnamefont {O.}~\bibnamefont
  {Pujol\`as}},\ }\href {\doibase 10.1103/PhysRevLett.114.251602} {\bibfield
  {journal} {\bibinfo  {journal} {Phys. Rev. Lett.}\ }\textbf {\bibinfo
  {volume} {114}},\ \bibinfo {pages} {251602} (\bibinfo {year}
  {2015})}\BibitemShut {NoStop}%
\bibitem [{\citenamefont {Alberte}\ \emph
  {et~al.}(2016{\natexlab{a}})\citenamefont {Alberte}, \citenamefont
  {Baggioli}, \citenamefont {Khmelnitsky},\ and\ \citenamefont
  {Pujolas}}]{Alberte:2015isw}%
  \BibitemOpen
  \bibfield  {author} {\bibinfo {author} {\bibfnamefont {L.}~\bibnamefont
  {Alberte}}, \bibinfo {author} {\bibfnamefont {M.}~\bibnamefont {Baggioli}},
  \bibinfo {author} {\bibfnamefont {A.}~\bibnamefont {Khmelnitsky}}, \ and\
  \bibinfo {author} {\bibfnamefont {O.}~\bibnamefont {Pujolas}},\ }\href
  {\doibase 10.1007/JHEP02(2016)114} {\bibfield  {journal} {\bibinfo  {journal}
  {JHEP}\ }\textbf {\bibinfo {volume} {02}},\ \bibinfo {pages} {114} (\bibinfo
  {year} {2016}{\natexlab{a}})},\ \Eprint {http://arxiv.org/abs/1510.09089}
  {arXiv:1510.09089 [hep-th]} \BibitemShut {NoStop}%
%%CITATION = ARXIV:1510.09089;%%
\bibitem [{\citenamefont {Alberte}\ \emph {et~al.}(2017)\citenamefont
  {Alberte}, \citenamefont {Ammon}, \citenamefont {Baggioli}, \citenamefont
  {Jiménez},\ and\ \citenamefont {Pujolàs}}]{Alberte:2017cch}%
  \BibitemOpen
  \bibfield  {author} {\bibinfo {author} {\bibfnamefont {L.}~\bibnamefont
  {Alberte}}, \bibinfo {author} {\bibfnamefont {M.}~\bibnamefont {Ammon}},
  \bibinfo {author} {\bibfnamefont {M.}~\bibnamefont {Baggioli}}, \bibinfo
  {author} {\bibfnamefont {A.}~\bibnamefont {Jiménez}}, \ and\ \bibinfo
  {author} {\bibfnamefont {O.}~\bibnamefont {Pujolàs}},\ }\href@noop {} {\
  (\bibinfo {year} {2017})},\ \Eprint {http://arxiv.org/abs/1708.08477}
  {arXiv:1708.08477 [hep-th]} \BibitemShut {NoStop}%
%%CITATION = ARXIV:1708.08477;%%
\bibitem [{\citenamefont {Landau}\ and\ \citenamefont
  {Lifshitz}(1970)}]{Landau7}%
  \BibitemOpen
  \bibfield  {author} {\bibinfo {author} {\bibfnamefont {L.~D.}\ \bibnamefont
  {Landau}}\ and\ \bibinfo {author} {\bibfnamefont {E.~M.}\ \bibnamefont
  {Lifshitz}},\ }\href@noop {} {\emph {\bibinfo {title} {Course of Theoretical
  Physics, Vol. 7,Theory of Elasticity}}}\ (\bibinfo  {publisher} {Pergamon
  Press},\ \bibinfo {year} {1970})\BibitemShut {NoStop}%
\bibitem [{Note1()}]{Note1}%
  \BibitemOpen
  \bibinfo {note} {The SSB of translations has been previously realized in
  holography in \cite
  {Nakamura:2009tf,Donos:2012wi,Donos:2013gda,Donos:2011bh,Andrade:2017cnc,Andrade:2017cnc,Jokela:2017ltu}
  as the dual of \protect \textit {charge density waves} states \cite
  {RevModPhys.60.1129} where the Goldstone boson is identified with the
  so-called \protect \textit {sliding mode}. Another realization of gapless
  transverse phonon modes was made in \cite {Esposito:2017qpj}, using a model
  with more dynamical ingredients -- even though a direct comparison to the
  elastic moduli is lacking.}\BibitemShut {Stop}%
\bibitem [{Note2()}]{Note2}%
  \BibitemOpen
  \bibinfo {note} {We thank Blaise Gouteraux for suggesting this
  interpretation. See also \cite {Amoretti:2016bxs} for a detailed analysis on
  the nature of the translational symmetry breaking patterns in QFT and
  holography.}\BibitemShut {Stop}%
\bibitem [{\citenamefont {Leutwyler}(1997)}]{Leutwyler:1996er}%
  \BibitemOpen
  \bibfield  {author} {\bibinfo {author} {\bibfnamefont {H.}~\bibnamefont
  {Leutwyler}},\ }\href@noop {} {\bibfield  {journal} {\bibinfo  {journal}
  {Helv. Phys. Acta}\ }\textbf {\bibinfo {volume} {70}},\ \bibinfo {pages}
  {275} (\bibinfo {year} {1997})},\ \Eprint
  {http://arxiv.org/abs/hep-ph/9609466} {arXiv:hep-ph/9609466 [hep-ph]}
  \BibitemShut {NoStop}%
%%CITATION = HEP-PH/9609466;%%
\bibitem [{\citenamefont {{Garai}}(2007)}]{2007physics...3001G}%
  \BibitemOpen
  \bibfield  {author} {\bibinfo {author} {\bibfnamefont {J.}~\bibnamefont
  {{Garai}}},\ }\href@noop {} {\bibfield  {journal} {\bibinfo  {journal} {ArXiv
  Physics e-prints}\ } (\bibinfo {year} {2007})},\ \Eprint
  {http://arxiv.org/abs/physics/0703001} {physics/0703001} \BibitemShut
  {NoStop}%
\bibitem [{Note3()}]{Note3}%
  \BibitemOpen
  \bibinfo {note} {This new parameter does not appear in models with a standard
  kinetic term near $X=0$, \protect \emph {i.e.}, $V(X)=X+X^2+\protect \dots $
  because there $\Lambda (u)$ does not asymptote to zero}\BibitemShut {NoStop}%
\bibitem [{\citenamefont {Nicolis}\ \emph {et~al.}(2015)\citenamefont
  {Nicolis}, \citenamefont {Penco}, \citenamefont {Piazza},\ and\ \citenamefont
  {Rattazzi}}]{Nicolis:2015sra}%
  \BibitemOpen
  \bibfield  {author} {\bibinfo {author} {\bibfnamefont {A.}~\bibnamefont
  {Nicolis}}, \bibinfo {author} {\bibfnamefont {R.}~\bibnamefont {Penco}},
  \bibinfo {author} {\bibfnamefont {F.}~\bibnamefont {Piazza}}, \ and\ \bibinfo
  {author} {\bibfnamefont {R.}~\bibnamefont {Rattazzi}},\ }\href {\doibase
  10.1007/JHEP06(2015)155} {\bibfield  {journal} {\bibinfo  {journal} {JHEP}\
  }\textbf {\bibinfo {volume} {06}},\ \bibinfo {pages} {155} (\bibinfo {year}
  {2015})},\ \Eprint {http://arxiv.org/abs/1501.03845} {arXiv:1501.03845
  [hep-th]} \BibitemShut {NoStop}%
%%CITATION = ARXIV:1501.03845;%%
\bibitem [{Note4()}]{Note4}%
  \BibitemOpen
  \bibinfo {note} {This is opposed to the case of mass potentials $V(X)=X^{n}$
  with $ n<5/2$, studied in \cite
  {Davison:2013jba,Davison:2014lua,Alberte:2017cch}.}\BibitemShut {Stop}%
\bibitem [{\citenamefont {Martin}\ \emph {et~al.}(1972)\citenamefont {Martin},
  \citenamefont {Parodi},\ and\ \citenamefont {Pershan}}]{PhysRevA.6.2401}%
  \BibitemOpen
  \bibfield  {author} {\bibinfo {author} {\bibfnamefont {P.~C.}\ \bibnamefont
  {Martin}}, \bibinfo {author} {\bibfnamefont {O.}~\bibnamefont {Parodi}}, \
  and\ \bibinfo {author} {\bibfnamefont {P.~S.}\ \bibnamefont {Pershan}},\
  }\href {\doibase 10.1103/PhysRevA.6.2401} {\bibfield  {journal} {\bibinfo
  {journal} {Phys. Rev. A}\ }\textbf {\bibinfo {volume} {6}},\ \bibinfo {pages}
  {2401} (\bibinfo {year} {1972})}\BibitemShut {NoStop}%
\bibitem [{\citenamefont {{Zippelius}}\ \emph {et~al.}(1980)\citenamefont
  {{Zippelius}}, \citenamefont {{Halperin}},\ and\ \citenamefont
  {{Nelson}}}]{1980PhRvB..22.2514Z}%
  \BibitemOpen
  \bibfield  {author} {\bibinfo {author} {\bibfnamefont {A.}~\bibnamefont
  {{Zippelius}}}, \bibinfo {author} {\bibfnamefont {B.~I.}\ \bibnamefont
  {{Halperin}}}, \ and\ \bibinfo {author} {\bibfnamefont {D.~R.}\ \bibnamefont
  {{Nelson}}},\ }\href {\doibase 10.1103/PhysRevB.22.2514} {\bibfield
  {journal} {\bibinfo  {journal} {\prb}\ }\textbf {\bibinfo {volume} {22}},\
  \bibinfo {pages} {2514} (\bibinfo {year} {1980})}\BibitemShut {NoStop}%
\bibitem [{\citenamefont {{Kadanoff}}\ and\ \citenamefont
  {{Martin}}(1963)}]{1963AnPhy..24..419K}%
  \BibitemOpen
  \bibfield  {author} {\bibinfo {author} {\bibfnamefont {L.~P.}\ \bibnamefont
  {{Kadanoff}}}\ and\ \bibinfo {author} {\bibfnamefont {P.~C.}\ \bibnamefont
  {{Martin}}},\ }\href {\doibase 10.1016/0003-4916(63)90078-2} {\bibfield
  {journal} {\bibinfo  {journal} {Annals of Physics}\ }\textbf {\bibinfo
  {volume} {24}},\ \bibinfo {pages} {419} (\bibinfo {year} {1963})}\BibitemShut
  {NoStop}%
\bibitem [{Note5()}]{Note5}%
  \BibitemOpen
  \bibinfo {note} {Due to the presence of a finite shear modulus $G$ the
  mechanical pressure, $\protect \mathcal {P}\equiv T_{xx}$, and the
  thermodynamic pressure, $p\equiv -\Omega /\protect \mathcal V$, are not
  equivalent.}\BibitemShut {Stop}%
\bibitem [{Note6()}]{Note6}%
  \BibitemOpen
  \bibinfo {note} {We have also checked the linear superpositions of the type
  $V(X)=\DOTSB \sum@ \slimits@ c_\nu X^\nu $ with $\nu >5/2$; the qualitatitve
  results remain unchanged.}\BibitemShut {Stop}%
\bibitem [{Note7()}]{Note7}%
  \BibitemOpen
  \bibinfo {note} {We fitted the data to $\omega = C_1k^{\alpha _1}+i\protect
  \tmspace +\thinmuskip {.1667em}C_2k^{\alpha _2}$. In all cases we found
  similar values: $\alpha _1=1.00001\pm 4\cdot 10^{-5}$ and $\alpha
  _2=2.00002\pm 1\cdot 10^{-5}$.}\BibitemShut {Stop}%
\bibitem [{\citenamefont {Hartnoll}\ \emph
  {et~al.}(2016{\natexlab{b}})\citenamefont {Hartnoll}, \citenamefont
  {Ramirez},\ and\ \citenamefont {Santos}}]{Hartnoll:2016tri}%
  \BibitemOpen
  \bibfield  {author} {\bibinfo {author} {\bibfnamefont {S.~A.}\ \bibnamefont
  {Hartnoll}}, \bibinfo {author} {\bibfnamefont {D.~M.}\ \bibnamefont
  {Ramirez}}, \ and\ \bibinfo {author} {\bibfnamefont {J.~E.}\ \bibnamefont
  {Santos}},\ }\href {\doibase 10.1007/JHEP03(2016)170} {\bibfield  {journal}
  {\bibinfo  {journal} {JHEP}\ }\textbf {\bibinfo {volume} {03}},\ \bibinfo
  {pages} {170} (\bibinfo {year} {2016}{\natexlab{b}})},\ \Eprint
  {http://arxiv.org/abs/1601.02757} {arXiv:1601.02757 [hep-th]} \BibitemShut
  {NoStop}%
%%CITATION = ARXIV:1601.02757;%%
\bibitem [{\citenamefont {Alberte}\ \emph
  {et~al.}(2016{\natexlab{b}})\citenamefont {Alberte}, \citenamefont
  {Baggioli},\ and\ \citenamefont {Pujolas}}]{Alberte:2016xja}%
  \BibitemOpen
  \bibfield  {author} {\bibinfo {author} {\bibfnamefont {L.}~\bibnamefont
  {Alberte}}, \bibinfo {author} {\bibfnamefont {M.}~\bibnamefont {Baggioli}}, \
  and\ \bibinfo {author} {\bibfnamefont {O.}~\bibnamefont {Pujolas}},\ }\href
  {\doibase 10.1007/JHEP07(2016)074} {\bibfield  {journal} {\bibinfo  {journal}
  {JHEP}\ }\textbf {\bibinfo {volume} {07}},\ \bibinfo {pages} {074} (\bibinfo
  {year} {2016}{\natexlab{b}})},\ \Eprint {http://arxiv.org/abs/1601.03384}
  {arXiv:1601.03384 [hep-th]} \BibitemShut {NoStop}%
%%CITATION = ARXIV:1601.03384;%%
\bibitem [{\citenamefont {Dyre}(2006)}]{RevModPhys.78.953}%
  \BibitemOpen
  \bibfield  {author} {\bibinfo {author} {\bibfnamefont {J.~C.}\ \bibnamefont
  {Dyre}},\ }\href {\doibase 10.1103/RevModPhys.78.953} {\bibfield  {journal}
  {\bibinfo  {journal} {Rev. Mod. Phys.}\ }\textbf {\bibinfo {volume} {78}},\
  \bibinfo {pages} {953} (\bibinfo {year} {2006})}\BibitemShut {NoStop}%
\bibitem [{\citenamefont {Berthier}\ and\ \citenamefont
  {Biroli}(2011)}]{RevModPhys.83.587}%
  \BibitemOpen
  \bibfield  {author} {\bibinfo {author} {\bibfnamefont {L.}~\bibnamefont
  {Berthier}}\ and\ \bibinfo {author} {\bibfnamefont {G.}~\bibnamefont
  {Biroli}},\ }\href {\doibase 10.1103/RevModPhys.83.587} {\bibfield  {journal}
  {\bibinfo  {journal} {Rev. Mod. Phys.}\ }\textbf {\bibinfo {volume} {83}},\
  \bibinfo {pages} {587} (\bibinfo {year} {2011})}\BibitemShut {NoStop}%
\bibitem [{\citenamefont {{Cavagna}}(2009)}]{2009PhR...476...51C}%
  \BibitemOpen
  \bibfield  {author} {\bibinfo {author} {\bibfnamefont {A.}~\bibnamefont
  {{Cavagna}}},\ }\href {\doibase 10.1016/j.physrep.2009.03.003} {\bibfield
  {journal} {\bibinfo  {journal} {Physics Reports}\ }\textbf {\bibinfo {volume}
  {476}},\ \bibinfo {pages} {51} (\bibinfo {year} {2009})},\ \Eprint
  {http://arxiv.org/abs/0903.4264} {arXiv:0903.4264 [cond-mat.stat-mech]}
  \BibitemShut {NoStop}%
\bibitem [{\citenamefont {Esposito}\ \emph {et~al.}(2017)\citenamefont
  {Esposito}, \citenamefont {Garcia-Saenz}, \citenamefont {Nicolis},\ and\
  \citenamefont {Penco}}]{Esposito:2017qpj}%
  \BibitemOpen
  \bibfield  {author} {\bibinfo {author} {\bibfnamefont {A.}~\bibnamefont
  {Esposito}}, \bibinfo {author} {\bibfnamefont {S.}~\bibnamefont
  {Garcia-Saenz}}, \bibinfo {author} {\bibfnamefont {A.}~\bibnamefont
  {Nicolis}}, \ and\ \bibinfo {author} {\bibfnamefont {R.}~\bibnamefont
  {Penco}},\ }\href@noop {} {\  (\bibinfo {year} {2017})},\ \Eprint
  {http://arxiv.org/abs/1708.09391} {arXiv:1708.09391 [hep-th]} \BibitemShut
  {NoStop}%
%%CITATION = ARXIV:1708.09391;%%
\bibitem [{\citenamefont {Davison}\ and\ \citenamefont
  {Goutéraux}(2015)}]{Davison:2014lua}%
  \BibitemOpen
  \bibfield  {author} {\bibinfo {author} {\bibfnamefont {R.~A.}\ \bibnamefont
  {Davison}}\ and\ \bibinfo {author} {\bibfnamefont {B.}~\bibnamefont
  {Goutéraux}},\ }\href {\doibase 10.1007/JHEP01(2015)039} {\bibfield
  {journal} {\bibinfo  {journal} {JHEP}\ }\textbf {\bibinfo {volume} {01}},\
  \bibinfo {pages} {039} (\bibinfo {year} {2015})},\ \Eprint
  {http://arxiv.org/abs/1411.1062} {arXiv:1411.1062 [hep-th]} \BibitemShut
  {NoStop}%
%%CITATION = ARXIV:1411.1062;%%
\bibitem [{\citenamefont {Davison}\ \emph {et~al.}(2015)\citenamefont
  {Davison}, \citenamefont {Goutéraux},\ and\ \citenamefont
  {Hartnoll}}]{Davison:2015taa}%
  \BibitemOpen
  \bibfield  {author} {\bibinfo {author} {\bibfnamefont {R.~A.}\ \bibnamefont
  {Davison}}, \bibinfo {author} {\bibfnamefont {B.}~\bibnamefont {Goutéraux}},
  \ and\ \bibinfo {author} {\bibfnamefont {S.~A.}\ \bibnamefont {Hartnoll}},\
  }\href {\doibase 10.1007/JHEP10(2015)112} {\bibfield  {journal} {\bibinfo
  {journal} {JHEP}\ }\textbf {\bibinfo {volume} {10}},\ \bibinfo {pages} {112}
  (\bibinfo {year} {2015})},\ \Eprint {http://arxiv.org/abs/1507.07137}
  {arXiv:1507.07137 [hep-th]} \BibitemShut {NoStop}%
%%CITATION = ARXIV:1507.07137;%%
\bibitem [{\citenamefont {Burikham}\ and\ \citenamefont
  {Poovuttikul}(2016)}]{Burikham:2016roo}%
  \BibitemOpen
  \bibfield  {author} {\bibinfo {author} {\bibfnamefont {P.}~\bibnamefont
  {Burikham}}\ and\ \bibinfo {author} {\bibfnamefont {N.}~\bibnamefont
  {Poovuttikul}},\ }\href {\doibase 10.1103/PhysRevD.94.106001} {\bibfield
  {journal} {\bibinfo  {journal} {Phys. Rev.}\ }\textbf {\bibinfo {volume}
  {D94}},\ \bibinfo {pages} {106001} (\bibinfo {year} {2016})},\ \Eprint
  {http://arxiv.org/abs/1601.04624} {arXiv:1601.04624 [hep-th]} \BibitemShut
  {NoStop}%
%%CITATION = ARXIV:1601.04624;%%
\bibitem [{\citenamefont {Ciobanu}\ and\ \citenamefont
  {Ramirez}(2017)}]{Ciobanu:2017fef}%
  \BibitemOpen
  \bibfield  {author} {\bibinfo {author} {\bibfnamefont {T.}~\bibnamefont
  {Ciobanu}}\ and\ \bibinfo {author} {\bibfnamefont {D.~M.}\ \bibnamefont
  {Ramirez}},\ }\href@noop {} {\  (\bibinfo {year} {2017})},\ \Eprint
  {http://arxiv.org/abs/1708.04997} {arXiv:1708.04997 [hep-th]} \BibitemShut
  {NoStop}%
%%CITATION = ARXIV:1708.04997;%%
\bibitem [{\citenamefont {Beekman}\ \emph {et~al.}(2017)\citenamefont
  {Beekman}, \citenamefont {Nissinen}, \citenamefont {Wu}, \citenamefont {Liu},
  \citenamefont {Slager}, \citenamefont {Nussinov}, \citenamefont {Cvetkovic},\
  and\ \citenamefont {Zaanen}}]{Beekman:2016szb}%
  \BibitemOpen
  \bibfield  {author} {\bibinfo {author} {\bibfnamefont {A.~J.}\ \bibnamefont
  {Beekman}}, \bibinfo {author} {\bibfnamefont {J.}~\bibnamefont {Nissinen}},
  \bibinfo {author} {\bibfnamefont {K.}~\bibnamefont {Wu}}, \bibinfo {author}
  {\bibfnamefont {K.}~\bibnamefont {Liu}}, \bibinfo {author} {\bibfnamefont
  {R.-J.}\ \bibnamefont {Slager}}, \bibinfo {author} {\bibfnamefont
  {Z.}~\bibnamefont {Nussinov}}, \bibinfo {author} {\bibfnamefont
  {V.}~\bibnamefont {Cvetkovic}}, \ and\ \bibinfo {author} {\bibfnamefont
  {J.}~\bibnamefont {Zaanen}},\ }\href {\doibase 10.1016/j.physrep.2017.03.004}
  {\bibfield  {journal} {\bibinfo  {journal} {Phys. Rept.}\ }\textbf {\bibinfo
  {volume} {683}},\ \bibinfo {pages} {1} (\bibinfo {year} {2017})},\ \Eprint
  {http://arxiv.org/abs/1603.04254} {arXiv:1603.04254 [cond-mat.str-el]}
  \BibitemShut {NoStop}%
%%CITATION = ARXIV:1603.04254;%%
\bibitem [{\citenamefont {Micoulaut}(2016)}]{0034-4885-79-6-066504}%
  \BibitemOpen
  \bibfield  {author} {\bibinfo {author} {\bibfnamefont {M.}~\bibnamefont
  {Micoulaut}},\ }\href {http://stacks.iop.org/0034-4885/79/i=6/a=066504}
  {\bibfield  {journal} {\bibinfo  {journal} {Reports on Progress in Physics}\
  }\textbf {\bibinfo {volume} {79}},\ \bibinfo {pages} {066504} (\bibinfo
  {year} {2016})}\BibitemShut {NoStop}%
\bibitem [{\citenamefont {Anninos}\ \emph {et~al.}(2015)\citenamefont
  {Anninos}, \citenamefont {Anous}, \citenamefont {Denef},\ and\ \citenamefont
  {Peeters}}]{Anninos:2013mfa}%
  \BibitemOpen
  \bibfield  {author} {\bibinfo {author} {\bibfnamefont {D.}~\bibnamefont
  {Anninos}}, \bibinfo {author} {\bibfnamefont {T.}~\bibnamefont {Anous}},
  \bibinfo {author} {\bibfnamefont {F.}~\bibnamefont {Denef}}, \ and\ \bibinfo
  {author} {\bibfnamefont {L.}~\bibnamefont {Peeters}},\ }\href {\doibase
  10.1007/JHEP04(2015)027} {\bibfield  {journal} {\bibinfo  {journal} {JHEP}\
  }\textbf {\bibinfo {volume} {04}},\ \bibinfo {pages} {027} (\bibinfo {year}
  {2015})},\ \Eprint {http://arxiv.org/abs/1309.0146} {arXiv:1309.0146
  [hep-th]} \BibitemShut {NoStop}%
%%CITATION = ARXIV:1309.0146;%%
\bibitem [{\citenamefont {{Zacharias}}\ \emph
  {et~al.}(2015{\natexlab{a}})\citenamefont {{Zacharias}}, \citenamefont
  {{Paul}},\ and\ \citenamefont {{Garst}}}]{2015PhRvL.115b5703Z}%
  \BibitemOpen
  \bibfield  {author} {\bibinfo {author} {\bibfnamefont {M.}~\bibnamefont
  {{Zacharias}}}, \bibinfo {author} {\bibfnamefont {I.}~\bibnamefont {{Paul}}},
  \ and\ \bibinfo {author} {\bibfnamefont {M.}~\bibnamefont {{Garst}}},\ }\href
  {\doibase 10.1103/PhysRevLett.115.025703} {\bibfield  {journal} {\bibinfo
  {journal} {Physical Review Letters}\ }\textbf {\bibinfo {volume} {115}},\
  \bibinfo {eid} {025703} (\bibinfo {year} {2015}{\natexlab{a}})},\ \Eprint
  {http://arxiv.org/abs/1411.6925} {arXiv:1411.6925 [cond-mat.str-el]}
  \BibitemShut {NoStop}%
\bibitem [{\citenamefont {{Zacharias}}\ \emph
  {et~al.}(2015{\natexlab{b}})\citenamefont {{Zacharias}}, \citenamefont
  {{Rosch}},\ and\ \citenamefont {{Garst}}}]{2015EPJST.224.1021Z}%
  \BibitemOpen
  \bibfield  {author} {\bibinfo {author} {\bibfnamefont {M.}~\bibnamefont
  {{Zacharias}}}, \bibinfo {author} {\bibfnamefont {A.}~\bibnamefont
  {{Rosch}}}, \ and\ \bibinfo {author} {\bibfnamefont {M.}~\bibnamefont
  {{Garst}}},\ }\href {\doibase 10.1140/epjst/e2015-02444-5} {\bibfield
  {journal} {\bibinfo  {journal} {European Physical Journal Special Topics}\
  }\textbf {\bibinfo {volume} {224}} (\bibinfo {year} {2015}{\natexlab{b}}),\
  10.1140/epjst/e2015-02444-5},\ \Eprint {http://arxiv.org/abs/1507.04157}
  {arXiv:1507.04157 [cond-mat.str-el]} \BibitemShut {NoStop}%
\bibitem [{\citenamefont {Delacrétaz}\ \emph {et~al.}(2016)\citenamefont
  {Delacrétaz}, \citenamefont {Goutéraux}, \citenamefont {Hartnoll},\ and\
  \citenamefont {Karlsson}}]{Delacretaz:2016ivq}%
  \BibitemOpen
  \bibfield  {author} {\bibinfo {author} {\bibfnamefont {L.~V.}\ \bibnamefont
  {Delacrétaz}}, \bibinfo {author} {\bibfnamefont {B.}~\bibnamefont
  {Goutéraux}}, \bibinfo {author} {\bibfnamefont {S.~A.}\ \bibnamefont
  {Hartnoll}}, \ and\ \bibinfo {author} {\bibfnamefont {A.}~\bibnamefont
  {Karlsson}},\ }\href {\doibase 10.21468/SciPostPhys.3.3.025} {\  (\bibinfo
  {year} {2016}),\ 10.21468/SciPostPhys.3.3.025},\ \Eprint
  {http://arxiv.org/abs/1612.04381} {arXiv:1612.04381 [cond-mat.str-el]}
  \BibitemShut {NoStop}%
%%CITATION = ARXIV:1612.04381;%%
\bibitem [{\citenamefont {Delacrétaz}\ \emph {et~al.}(2017)\citenamefont
  {Delacrétaz}, \citenamefont {Goutéraux}, \citenamefont {Hartnoll},\ and\
  \citenamefont {Karlsson}}]{Delacretaz:2017zxd}%
  \BibitemOpen
  \bibfield  {author} {\bibinfo {author} {\bibfnamefont {L.~V.}\ \bibnamefont
  {Delacrétaz}}, \bibinfo {author} {\bibfnamefont {B.}~\bibnamefont
  {Goutéraux}}, \bibinfo {author} {\bibfnamefont {S.~A.}\ \bibnamefont
  {Hartnoll}}, \ and\ \bibinfo {author} {\bibfnamefont {A.}~\bibnamefont
  {Karlsson}},\ }\href@noop {} {\  (\bibinfo {year} {2017})},\ \Eprint
  {http://arxiv.org/abs/1702.05104} {arXiv:1702.05104 [cond-mat.str-el]}
  \BibitemShut {NoStop}%
%%CITATION = ARXIV:1702.05104;%%
\bibitem [{\citenamefont {Grozdanov}\ and\ \citenamefont
  {Poovuttikul}()}]{NickAppear}%
  \BibitemOpen
  \bibfield  {author} {\bibinfo {author} {\bibfnamefont {S.}~\bibnamefont
  {Grozdanov}}\ and\ \bibinfo {author} {\bibfnamefont {N.}~\bibnamefont
  {Poovuttikul}},\ }\href@noop {} {\bibinfo  {journal} {to appear}\
  }\BibitemShut {NoStop}%
\bibitem [{\citenamefont {Amoretti}\ \emph {et~al.}()\citenamefont {Amoretti},
  \citenamefont {Arean}, \citenamefont {Goutéraux},\ and\ \citenamefont
  {Musso}}]{BlaiseAppear}%
  \BibitemOpen
\bibfield  {journal} {  }\bibfield  {author} {\bibinfo {author} {\bibfnamefont
  {A.}~\bibnamefont {Amoretti}}, \bibinfo {author} {\bibfnamefont
  {D.}~\bibnamefont {Arean}}, \bibinfo {author} {\bibfnamefont
  {B.}~\bibnamefont {Goutéraux}}, \ and\ \bibinfo {author} {\bibfnamefont
  {D.}~\bibnamefont {Musso}},\ }\href@noop {} {\bibinfo  {journal} {to appear}\
  }\BibitemShut {NoStop}%
\bibitem [{\citenamefont {Nakamura}\ \emph {et~al.}(2010)\citenamefont
  {Nakamura}, \citenamefont {Ooguri},\ and\ \citenamefont
  {Park}}]{Nakamura:2009tf}%
  \BibitemOpen
\bibfield  {journal} {  }\bibfield  {author} {\bibinfo {author} {\bibfnamefont
  {S.}~\bibnamefont {Nakamura}}, \bibinfo {author} {\bibfnamefont
  {H.}~\bibnamefont {Ooguri}}, \ and\ \bibinfo {author} {\bibfnamefont {C.-S.}\
  \bibnamefont {Park}},\ }\href {\doibase 10.1103/PhysRevD.81.044018}
  {\bibfield  {journal} {\bibinfo  {journal} {Phys. Rev.}\ }\textbf {\bibinfo
  {volume} {D81}},\ \bibinfo {pages} {044018} (\bibinfo {year} {2010})},\
  \Eprint {http://arxiv.org/abs/0911.0679} {arXiv:0911.0679 [hep-th]}
  \BibitemShut {NoStop}%
%%CITATION = ARXIV:0911.0679;%%
\bibitem [{\citenamefont {Donos}\ and\ \citenamefont
  {Gauntlett}(2012)}]{Donos:2012wi}%
  \BibitemOpen
  \bibfield  {author} {\bibinfo {author} {\bibfnamefont {A.}~\bibnamefont
  {Donos}}\ and\ \bibinfo {author} {\bibfnamefont {J.~P.}\ \bibnamefont
  {Gauntlett}},\ }\href {\doibase 10.1103/PhysRevD.86.064010} {\bibfield
  {journal} {\bibinfo  {journal} {Phys. Rev.}\ }\textbf {\bibinfo {volume}
  {D86}},\ \bibinfo {pages} {064010} (\bibinfo {year} {2012})},\ \Eprint
  {http://arxiv.org/abs/1204.1734} {arXiv:1204.1734 [hep-th]} \BibitemShut
  {NoStop}%
%%CITATION = ARXIV:1204.1734;%%
\bibitem [{\citenamefont {Donos}\ and\ \citenamefont
  {Gauntlett}(2013)}]{Donos:2013gda}%
  \BibitemOpen
  \bibfield  {author} {\bibinfo {author} {\bibfnamefont {A.}~\bibnamefont
  {Donos}}\ and\ \bibinfo {author} {\bibfnamefont {J.~P.}\ \bibnamefont
  {Gauntlett}},\ }\href {\doibase 10.1103/PhysRevD.87.126008} {\bibfield
  {journal} {\bibinfo  {journal} {Phys. Rev.}\ }\textbf {\bibinfo {volume}
  {D87}},\ \bibinfo {pages} {126008} (\bibinfo {year} {2013})},\ \Eprint
  {http://arxiv.org/abs/1303.4398} {arXiv:1303.4398 [hep-th]} \BibitemShut
  {NoStop}%
%%CITATION = ARXIV:1303.4398;%%
\bibitem [{\citenamefont {Donos}\ and\ \citenamefont
  {Gauntlett}(2011)}]{Donos:2011bh}%
  \BibitemOpen
  \bibfield  {author} {\bibinfo {author} {\bibfnamefont {A.}~\bibnamefont
  {Donos}}\ and\ \bibinfo {author} {\bibfnamefont {J.~P.}\ \bibnamefont
  {Gauntlett}},\ }\href {\doibase 10.1007/JHEP08(2011)140} {\bibfield
  {journal} {\bibinfo  {journal} {JHEP}\ }\textbf {\bibinfo {volume} {08}},\
  \bibinfo {pages} {140} (\bibinfo {year} {2011})},\ \Eprint
  {http://arxiv.org/abs/1106.2004} {arXiv:1106.2004 [hep-th]} \BibitemShut
  {NoStop}%
%%CITATION = ARXIV:1106.2004;%%
\bibitem [{\citenamefont {Andrade}\ \emph {et~al.}(2017)\citenamefont
  {Andrade}, \citenamefont {Baggioli}, \citenamefont {Krikun},\ and\
  \citenamefont {Poovuttikul}}]{Andrade:2017cnc}%
  \BibitemOpen
  \bibfield  {author} {\bibinfo {author} {\bibfnamefont {T.}~\bibnamefont
  {Andrade}}, \bibinfo {author} {\bibfnamefont {M.}~\bibnamefont {Baggioli}},
  \bibinfo {author} {\bibfnamefont {A.}~\bibnamefont {Krikun}}, \ and\ \bibinfo
  {author} {\bibfnamefont {N.}~\bibnamefont {Poovuttikul}},\ }\href@noop {} {\
  (\bibinfo {year} {2017})},\ \Eprint {http://arxiv.org/abs/1708.08306}
  {arXiv:1708.08306 [hep-th]} \BibitemShut {NoStop}%
%%CITATION = ARXIV:1708.08306;%%
\bibitem [{\citenamefont {Jokela}\ \emph {et~al.}(2017)\citenamefont {Jokela},
  \citenamefont {Jarvinen},\ and\ \citenamefont {Lippert}}]{Jokela:2017ltu}%
  \BibitemOpen
  \bibfield  {author} {\bibinfo {author} {\bibfnamefont {N.}~\bibnamefont
  {Jokela}}, \bibinfo {author} {\bibfnamefont {M.}~\bibnamefont {Jarvinen}}, \
  and\ \bibinfo {author} {\bibfnamefont {M.}~\bibnamefont {Lippert}},\
  }\href@noop {} {\  (\bibinfo {year} {2017})},\ \Eprint
  {http://arxiv.org/abs/1708.07837} {arXiv:1708.07837 [hep-th]} \BibitemShut
  {NoStop}%
%%CITATION = ARXIV:1708.07837;%%
\bibitem [{\citenamefont {Gr\"uner}(1988)}]{RevModPhys.60.1129}%
  \BibitemOpen
  \bibfield  {author} {\bibinfo {author} {\bibfnamefont {G.}~\bibnamefont
  {Gr\"uner}},\ }\href {\doibase 10.1103/RevModPhys.60.1129} {\bibfield
  {journal} {\bibinfo  {journal} {Rev. Mod. Phys.}\ }\textbf {\bibinfo {volume}
  {60}},\ \bibinfo {pages} {1129} (\bibinfo {year} {1988})}\BibitemShut
  {NoStop}%
\bibitem [{\citenamefont {Amoretti}\ \emph {et~al.}(2017)\citenamefont
  {Amoretti}, \citenamefont {Areán}, \citenamefont {Argurio}, \citenamefont
  {Musso},\ and\ \citenamefont {Pando~Zayas}}]{Amoretti:2016bxs}%
  \BibitemOpen
  \bibfield  {author} {\bibinfo {author} {\bibfnamefont {A.}~\bibnamefont
  {Amoretti}}, \bibinfo {author} {\bibfnamefont {D.}~\bibnamefont {Areán}},
  \bibinfo {author} {\bibfnamefont {R.}~\bibnamefont {Argurio}}, \bibinfo
  {author} {\bibfnamefont {D.}~\bibnamefont {Musso}}, \ and\ \bibinfo {author}
  {\bibfnamefont {L.~A.}\ \bibnamefont {Pando~Zayas}},\ }\href {\doibase
  10.1007/JHEP05(2017)051} {\bibfield  {journal} {\bibinfo  {journal} {JHEP}\
  }\textbf {\bibinfo {volume} {05}},\ \bibinfo {pages} {051} (\bibinfo {year}
  {2017})},\ \Eprint {http://arxiv.org/abs/1611.09344} {arXiv:1611.09344
  [hep-th]} \BibitemShut {NoStop}%
%%CITATION = ARXIV:1611.09344;%%
\bibitem [{\citenamefont {Baggioli}\ and\ \citenamefont
  {Brattan}(2017)}]{Baggioli:2015gsa}%
  \BibitemOpen
  \bibfield  {author} {\bibinfo {author} {\bibfnamefont {M.}~\bibnamefont
  {Baggioli}}\ and\ \bibinfo {author} {\bibfnamefont {D.~K.}\ \bibnamefont
  {Brattan}},\ }\href {\doibase 10.1088/1361-6382/34/1/015008} {\bibfield
  {journal} {\bibinfo  {journal} {Class. Quant. Grav.}\ }\textbf {\bibinfo
  {volume} {34}},\ \bibinfo {pages} {015008} (\bibinfo {year} {2017})},\
  \Eprint {http://arxiv.org/abs/1504.07635} {arXiv:1504.07635 [hep-th]}
  \BibitemShut {NoStop}%
%%CITATION = ARXIV:1504.07635;%%
\bibitem [{\citenamefont {Ammon}\ \emph {et~al.}(2017)\citenamefont {Ammon},
  \citenamefont {Kaminski}, \citenamefont {Koirala}, \citenamefont {Leiber},\
  and\ \citenamefont {Wu}}]{Ammon:2017ded}%
  \BibitemOpen
  \bibfield  {author} {\bibinfo {author} {\bibfnamefont {M.}~\bibnamefont
  {Ammon}}, \bibinfo {author} {\bibfnamefont {M.}~\bibnamefont {Kaminski}},
  \bibinfo {author} {\bibfnamefont {R.}~\bibnamefont {Koirala}}, \bibinfo
  {author} {\bibfnamefont {J.}~\bibnamefont {Leiber}}, \ and\ \bibinfo {author}
  {\bibfnamefont {J.}~\bibnamefont {Wu}},\ }\href {\doibase
  10.1007/JHEP04(2017)067} {\bibfield  {journal} {\bibinfo  {journal} {JHEP}\
  }\textbf {\bibinfo {volume} {04}},\ \bibinfo {pages} {067} (\bibinfo {year}
  {2017})},\ \Eprint {http://arxiv.org/abs/1701.05565} {arXiv:1701.05565
  [hep-th]} \BibitemShut {NoStop}%
%%CITATION = ARXIV:1701.05565;%%
\end{thebibliography}%

\appendix
\section{Supplementary material}\label{SUPPL}
\noindent In this section we present the details of the holographic model. From now on we set the AdS radius $\ell=1$ and $M_P=1$.\\

\noindent\textbf{\small{Background and thermodynamics}}.\\[0.15cm]
 As explained in the main text, the choice $\phi^I=x^I$ for the background of the scalar fields leads to a consistent, homogeneous solution of the equations of motion. In order to introduce also a charge density in the system, a non-vanishing gauge field is needed. The latter satisfies $A =\mu (1-u/u_h)\,dt$ where $\mu$ is the chemical potential. The solution for the emblackening factor $f$ for generic potentials $V(X)$ is then given by
\begin{equation}\label{backf}
f(u)= u^3 \int_u^{u_h} dv\;\left[ \frac{3}{v^4} -\frac{\mu^2}{2 \, u_h^2}-\frac{m^2}{v^4}\, 
V(v^2) \right] \, ,
\end{equation}      
where $u_h$ stands for the location of the black brane horizon. The temperature reads
\begin{equation}
T=-\frac{f'(u_h)}{4\pi}=\frac{6 - {\mu^2 u_h^2} -  2 m^2 V\left(u_h^2 \right) }{8 \pi u_h}~.
\end{equation}\\
For potentials of the form $V(X)=X^n$, the energy density of the black brane reads
\begin{equation}
\epsilon =\frac{1}{u_h^3}\left[1+\frac{\mu^2 u_h^2}{2}+\frac{m^2 \,  u_h^{2n}}{2n-3}\right]  \, 
\end{equation}
and can be easily derived from \eqref{backf}. The grand canonical potential for these models is
\begin{equation}\label{freenergy}
    \Omega =-\frac{\mathcal{V}}{2}\left[\frac{1}{u_h^3}+\frac{\mu}{2u_h^2}-\frac{2n-1}{2n-3} m^2 u_h^{2n-3}\right]\,,
\end{equation}
where $\mathcal{V}$ is the area of the spatial boundary. %{The thermodynamic pressure density is defined by the free energy as $p=-\Omega/V$.
We note that only in the regime $n\in \left[\frac{1}{2},\frac{3}{2}\right]$ the contribution to the free energy due to the mass term is negative. For other powers of $n$, it is the scalar field solution $\phi^I=0$ that defines the minimum of the potential. This raises a natural concern that 
%(in case one considers the ensemble of solutions that allows the  $\phi^I=0$ solution) 
for these values of $n$ the thermodynamic ground state should be determined by the $\phi^I=0$ solution, and not the solution $\phi^I=x^I$. The resolution to this concern is that, as argued in the main text, the $\phi^I=0$ solution is in the strong coupling regime and therefore even its existence as a saddle point is questionable.\\[0.15cm]
\noindent\textbf{\small{Heat capacity and elasticity}}.\\[0.15cm]
\begin{figure}[htp]
\begin{center}
\includegraphics[width=0.48\textwidth]{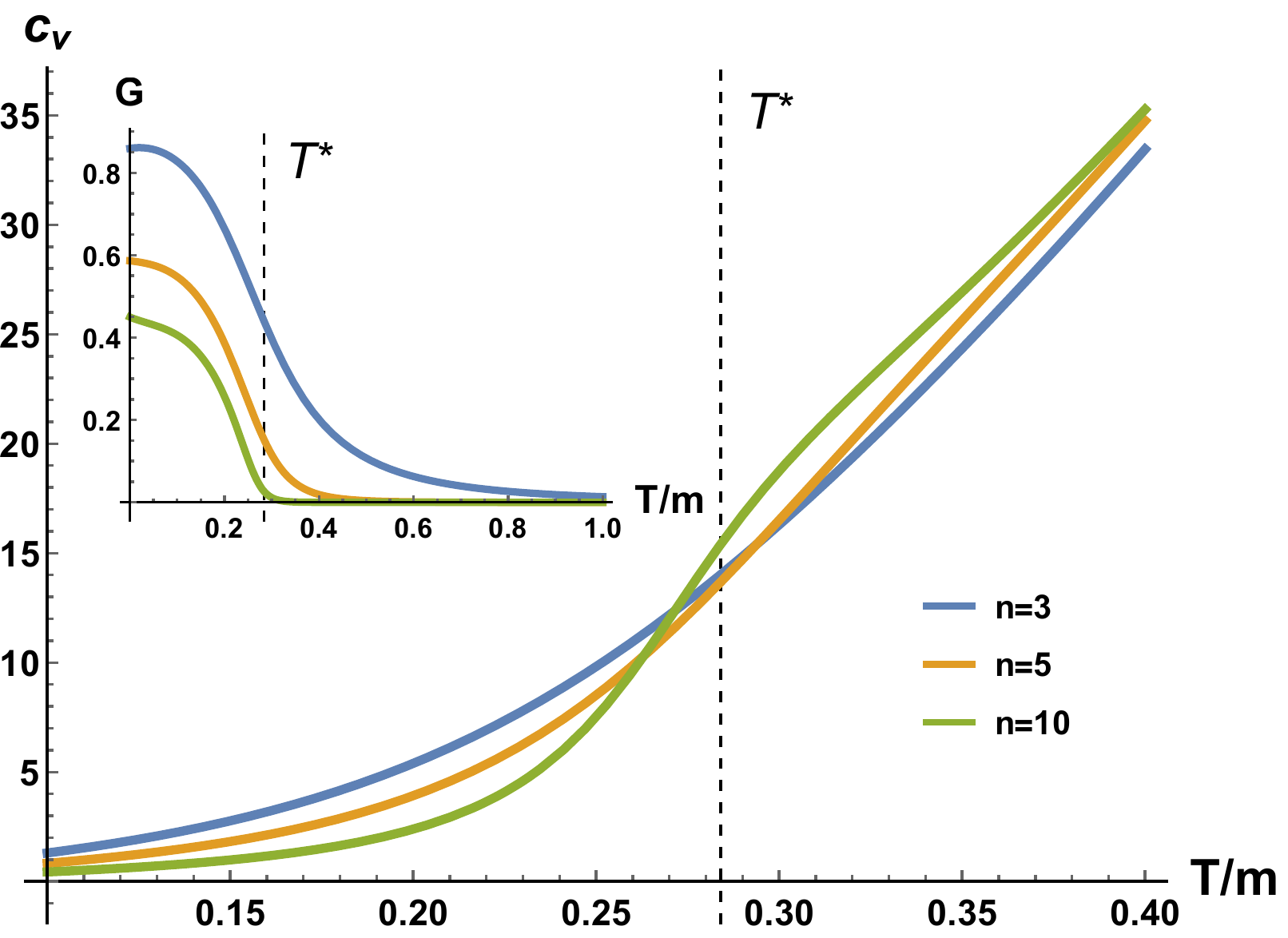}%&
 \caption{The heat capacity for $V(X)=X^n,$ with $n=3,5,10$ as a function of $T/m$. The inset shows the corresponding behaviour of the elastic shear modulus. The gray lines guide the eye towards the correlation between the two quantitities at the crossover scale $T^*$ for the case $n=10$.}
 \label{fig:heat}
\end{center}
\end{figure}
Given the expression for the entropy density in our setup, $s=2 \pi/u_h^2$, we can easily compute, as already analyzed in  \cite{Baggioli:2015gsa}, the heat capacity at zero chemical potential as:
\begin{equation}
c_v\,=\,T\,\frac{ds}{dT}\,=\,T\,\frac{ds}{d u_h}\left(\frac{dT}{d u_h}\right)^{-1}\,.
\end{equation}
We obtain:
\begin{equation}
c_v\,=\,\frac{4 \pi  \left(3-m^2\, V\right)}{u_h^2 \left(2\, m^2\, u_h^2 V'-m^2\, V+3\right)}
\end{equation}
where the potential $V$ is computed on the background value evaluated at the horizon $\equiv u_h^2$.

The case of $V(X)=X^n$ is shown in Fig.~\ref{fig:heat}. We see that the heat capacity exhibits a crossover at some specific temperature $T^*$, which becomes sharper with increasing $n$. There appears to be a correlation between the crossover in the heat capacity and in the shear modulus (shown in the inset of Fig.~\ref{fig:heat}): both occur at approximately the same temperature. We further note that the crossover temperature $T^*$ decreases with increasing $n$ in both cases. At least qualitatively, this is intriguingly similar to the properties of glass transitions \cite{RevModPhys.78.953,RevModPhys.83.587,2009PhR...476...51C}. In particular, it is compelling to identify the crossover scale $T^*$ with the glass transition temperature $T_g$. However, a more in-depth analysis is certainly needed.

The elastic shear modulus defined in the main text and shown in Fig.~\ref{fig:comparison} and Fig.~\ref{fig:heat} can be computed analytically for $m/T\ll 1$ as in \cite{Alberte:2016xja}. In particular, we obtain:
\begin{equation}
G\,=\,m^2\,\int_0^{u_h}\,\frac{V'\left(v^2\right)}{v^2}\,dv\,,\qquad m/T \ll 1\,.
\end{equation}
which is convergent for the cases of interest with $n>3/2$. For our choice of potentials \eqref{Xn}, the above expression becomes:
\begin{equation}\label{cc}
G\,=\,\frac{n}{2\,n\,-\,3}\,u_h^{2n-3}\,m^2\,,
\end{equation}
and shows a very good agreement with the numerical solution in the limit of $m/T\ll 1$.
The latter also predicts the high temperature fall-off of the shear modulus as $G\sim m^2T^{3-2n}$, shown in the inset of Fig.~\ref{fig:comparison}.\\[0.1cm]

\noindent \textbf{\small{Perturbations and Green's functions.}}\\[0.15cm]
Next we study the fluctuations on top of the background. We choose the momentum to be parallel to the $y$-axis. The transverse perturbations are then encoded in the fluctuations $a_x,\,h_{tx}\equiv u^2 \delta g_{tx},\,h_{xy}\equiv u^2\delta g_{xy},\,\delta \phi_x,\,\delta g_{xu}$. Assuming for simplicity the radial gauge, \emph{i.e.} $\delta g_{xu}=0$, and using the ingoing Eddington-Finkelstein coordinates, $ds^2=\frac{1}{u^2} \left[-f(u)dt^2-2dtdu + dx^2+dy^2\right]$,
the remaining equations read:
\begin{align}
&0=-2(1-u^2\,V''/V')h_{tx}+u\,h_{tx}'-i\,k\,u\, h_{xy}
\nonumber\\&-\left( k^2\,u+2\,i\,\omega (1-u^2\,V''/V')\right)\,\delta \phi_x+u\,f\, \delta \phi_x''
\nonumber\\&+\left(-2(1-u^2\,V''/V')\,f+u\,(2 i \omega+f')\right)\delta \phi_x'
\,\,;\nonumber\\
&0=2\,i\,m^2\,u^{2}\omega V'\,\delta \phi_x+u^2\, k\,\omega\, h_{xy}+2\,u^4\,\mu\,(i\omega\,a_x+f\,a_x')
\nonumber\\&+\left(6+k^2\,u^2-4\,u^2\mu^2-2 \, m^2 (V-u^2\,V' ) \, -6 f+2uf'\right) \, h_{tx}
\nonumber\\&+\left(2\,u\,f-i\,u^2\omega\right)h_{tx}'-u^2\,f\,h_{tx}'' \,\,;
\nonumber\\
&0=2i\,k\,u\,h_{tx}-iku^2h_{tx}'-2\,i \, k \,m^2 \,u^{2}V'\delta \phi_x
\nonumber\\&+2 h_{xy}\left(3+i\,u \, \omega-3f+uf'- \,m^2(V-u^2 V') \right)
\nonumber\\&-\left(2i\,u^2 \, \omega-2uf+u^2\,f'\right)h_{xy}'-u^2\,f\, h_{xy}''\,\,;
\nonumber\\
&0=2\,h_{tx}'-u\,h_{tx}''-2m^2\,u\,V' \, \delta\phi_x'+ik\,u\,h_{xy}'+2u^4 \,\mu \, a_x'\,\,;\nonumber\\
&0=-\mu\,h_{tx}'-k^2\,a_x+(2i\,\omega+f')\,a_x'+f\,a_x'' \,.
\end{align}
The asymptotics of the various bulk fields close to the UV boundary $u=0$ are:
\begin{align}
&\delta \phi_x\,=\,\phi_{x\,(l)}\,(1\,+\,\dots)+\,\phi_{x\,(s)}\,u^{5-2n}\,(1\,+\,\dots)\,,\quad\nonumber\\ &h_{tx}\,=h_{tx\,(l)}\,(1\,+\,\dots)\,+\,h_{tx\,(s)}\,u^{3}\,(1\,+\,\dots)\,,\nonumber\\ &h_{xy}\,=h_{xy\,(l)}\,(1\,+\,\dots)\,+\,h_{xy\,(s)}\,u^{3}\,(1\,+\,\dots)\,,\nonumber\\\quad &a_x\,=a_{x\,(l)}\,(1\,+\,\dots)\,+\,a_{x\,(s)}\,u\,(1\,+\,\dots)\,\,\,.
\end{align}
In these coordinates the ingoing boundary conditions at the horizon are automatically satisfied by regular solutions.
It follows that the various retarded Green's functions are defined as:
\begin{align}\label{greenF}
&\mathcal{G}^{\textrm{(R)}}_{T_{tx}T_{tx}}\,=\,\frac{2\,\Delta-d}{2}\,\frac{h_{tx\,(s)}}{h_{tx\,(l)}}\,=\,\frac{3}{2}\frac{h_{tx\,(s)}}{h_{tx\,(l)}}\,,\nonumber\\
&\mathcal{G}^{\textrm{(R)}}_{T_{xy}T_{xy}}\,=\,\frac{2\,\Delta-d}{2}\,\frac{h_{xy\,(s)}}{h_{xy\,(l)}}\,=\,\frac{3}{2}\frac{h_{xy\,(s)}}{h_{xy\,(l)}}\,,\nonumber\\
&\mathcal{G}^{\textrm{(R)}}_{JJ}\,=\,\frac{a_{x\,(s)}}{a_{x\,(l)}}\,,
\end{align}
where the time and spacetime dependences are omitted for simplicity. All the physical observables analyzed in this letter can be extracted from the retarded Green's functions. The QNMs coincide with the poles of the correlators given above which are computed numerically at finite frequency $\omega$ and momentum $k$ along the lines of~\cite{Ammon:2017ded}.

\end{document}